\documentclass[12pt, a4paper]{amsart}
\usepackage{amsmath, amsfonts, amsthm, amssymb, bm, bbm}
\usepackage{graphicx}
\usepackage{enumerate}
\usepackage{enumitem}
\usepackage{multirow}
\usepackage{color}
\usepackage{natbib}
\usepackage[colorlinks,citecolor=blue,urlcolor=blue]{hyperref}

\usepackage{amsaddr}

\newcommand{\logit}{\text{logit}}

\newcommand{\Be}{\text{Beta}}

\newcommand{\Ga}{\text{Ga}}
\newcommand{\IG}{\text{Inv-Ga}}
\newcommand{\GP}{\text{GP}}
\newcommand{\Bin}{\text{Bin}}

\newcommand{\N}{\text{N}}
\newcommand{\E}{\text{E}}

\newcommand{\bx}{\bm{x}}
\newcommand{\by}{\bm{y}}
\newcommand{\bbeta}{\bm{\beta}}
\newcommand{\tbeta}{\tilde{\beta}}

\newcommand{\tgamma}{\tilde{\gamma}}
\newcommand{\bmu}{\bm{\mu}}
\newcommand{\bmeta}{\bm{\eta}}
\newcommand{\btheta}{\bm{\theta}}

\newcommand{\bB}{\bm{B}}

\newcommand{\bV}{\bm{V}}
\newcommand{\tr}{\text{tr}}
\newcommand{\te}{\text{te}}

\makeatletter
\newcommand{\addresseshere}{%
  \enddoc@text\let\enddoc@text\relax
}
\makeatother

\makeatletter
\def\paragraph{\@startsection{paragraph}{4}%
  \z@\z@{-\fontdimen2\font}%
  {\normalfont\itshape}}
\makeatother

\definecolor{brown}{rgb}{0.8, 0.33, 0.1}

\begin{document}

\def\spacingset#1{\renewcommand{\baselinestretch}%
{#1}\small\normalsize} \spacingset{1}


\title[Bayesian Inference for the Transmission Dynamics of COVID-19]{Semiparametric Bayesian Inference for the Transmission Dynamics of COVID-19 with a State-Space Model}

\author[T. Zhou and Y. Ji]{Tianjian Zhou$^1$ and Yuan Ji$^2$}
\address{Department of Public Health Sciences, University of Chicago}
\email{$^1$tjzhou@uchicago.edu, $^2$yji@health.bsd.uchicago.edu}

\keywords{Effective reproduction number; Forecasting; Gaussian process; Infectious disease; Parallel tempering; SIR model}

\begin{abstract}
The outbreak of Coronavirus Disease 2019 (COVID-19) is an ongoing pandemic affecting over 200 countries and regions.
Inference about the transmission dynamics of COVID-19 can provide important insights into the speed of disease spread and the effects of mitigation policies.
We develop a novel Bayesian approach to such inference based on a probabilistic compartmental model using data of daily confirmed COVID-19 cases.
In particular, we consider a probabilistic extension of the classical susceptible-infectious-recovered model, which takes into account undocumented infections and allows the epidemiological parameters to vary over time.
We estimate the disease transmission rate via a Gaussian process prior, which captures nonlinear changes over time without the need of specific parametric assumptions.
We utilize a parallel-tempering Markov chain Monte Carlo algorithm to efficiently sample from the highly correlated posterior space. 
Predictions for future observations are done by sampling from their posterior predictive distributions.
Performance of the proposed approach is assessed using simulated datasets.
Finally, our approach is applied to COVID-19 data from six states of the United States: Washington, New York, California, Florida, Texas, and Illinois.
An R package \texttt{BaySIR} is made available at \url{https://github.com/tianjianzhou/BaySIR} for the public to conduct independent analysis or reproduce the results in this paper.
\end{abstract}


\spacingset{1.45}

\maketitle

\section{Introduction}
\label{sec:intro}

The outbreak of Coronavirus Disease 2019 (COVID-19), caused by Severe Acute Respiratory Syndrome Coronavirus 2 (SARS‑CoV‑2), was declared a pandemic on March 11, 2020 by the World Health Organization. 
As of July 2, 2020, the number of confirmed COVID-19 cases worldwide has exceeded 10.6 million, and the death toll has surpassed 516,000.
In order to control the spread of the virus, countries around the world have implemented unprecedented non-pharmaceutical interventions, such as case isolation, closure of schools, stay-at-home orders, banning of mass gatherings, and local and national lockdowns.
At the same time, social distancing and mask wearing by the public also contribute to the containment of COVID-19.

Researchers have made substantial efforts to study the transmission dynamics of COVID-19, evaluate the effects of government interventions, and forecast infection and death counts.
These works include \cite{aguilar2020investigating, chen2020scenario, flaxman2020estimating, giordano2020modelling, gomez2020infekta, gu2020better, ihme2020forecasting, li2020early, li2020substantial, pan2020association, sun2020tracking, wang2020spatiotemporal,  wang2020epidemiological, woody2020projections, wu2020nowcasting, zhang2020prediction},
among many others.
The modeling approaches taken by these works can be broadly categorized into three groups: (i) curve fitting, (ii) compartmental modeling, and (iii) agent-based modeling.
Curve fitting approaches fit a curve to the observed number of confirmed cases or deaths. For example, \cite{ihme2020forecasting} use a Gaussian error function to model the cumulative death rate at a specific location.
Compartmental modeling approaches
(e.g., \citealp{li2020substantial}) consider a partition of the population into compartments corresponding to different stages of the disease, and characterize the transmission dynamics of the disease by the flow of individuals through compartments.
Finally, agent-based modeling approaches (e.g., \citealp{gomez2020infekta}) use computer simulations to study the dynamic interactions among the agents (e.g., people in epidemiology) and between an agent and the environment.

In this paper, we develop a novel semiparametric Bayesian approach to modeling the transmission dynamics of COVID-19, which is critical for characterizing disease spread.
We aim to address a few issues related to the COVID-19 pandemic.
First, we provide estimation of key epidemiological parameters, such as the effective reproduction number of COVID-19. 
The Bayesian framework allows us to elicit informative priors for parameters that are difficult to estimate due to lack of data based on clinical characteristics of COVID-19, and also offers coherent uncertainty quantification for the parameter estimates.
Our second goal is to make predictions about the future trends of the spread of COVID-19 (e.g., future case counts), which will be done by calculating the posterior predictive distributions for the future observations.
Although such predictions are technically straightforward, we avoid overinterpretation of the predictions because 
they rely on extrapolation of highly unpredictable human behaviors and the number of diagnostic tests that will be deployed. 
Nevertheless, such predictions may be useful for the public and decision makers to understand the trends and future impacts of COVID-19 based on current rates of transmission. We shall see this in our case studies later.
Our analysis will be based on a probabilistic compartmental model motivated by the classical susceptible-infectious-recovered model \citep{kermack1927contribution}.
We will use data of daily confirmed COVID-19 cases reported by the Center for Systems Science and Engineering at Johns Hopkins University (JHU CSSE) \citep{dong2020interactive}.
We provide an R package \texttt{BaySIR}, available at \url{https://github.com/tianjianzhou/BaySIR}, that can be used to conduct independent analysis of COVID-19 data or reproduce the results in this paper.

The proposed Bayesian approach attempts to improve COVID-19 modeling in at least four aspects.
First, we explicitly model the number of \emph{undocumented infections}, which is  only considered by some, but not all, existing works. 
Due to the potentially limited testing capacity and the existence of pre-symptomatic and asymptomatic COVID-19 cases \citep{rothe2020transmission, he2020temporal}, many infected individuals may not have been detected as having the disease. Therefore, modeling of undocumented infections is essential for accurate inference.
Second, we estimate the disease transmission rate via Gaussian process regression (GPR), a semiparametric regression method.
The GPR approach is highly flexible and captures nonlinear and non-monotonic relationships without the need of specific parametric assumptions.
Third, we develop a parallel-tempering Markov chain Monte Carlo (PTMCMC) algorithm to efficiently sample from the posterior distribution of the epidemiological parameters, which leads to improvements in convergence and mixing compared to a standard MCMC procedure.
We find that standard MCMC cannot produce reliable inference due to poor mixing.
Lastly, we rigorously assess our approach through simulation studies, sensitivity analyses, cross-validation and goodness-of-fit tests. Such validations provide insights into the modeling of COVID-19 data, not only for our approach, but also for others based on similar assumptions such as the popular compartmental modeling approaches.

The remainder of the paper is organized as follows.
In Section \ref{sec:review_sir}, we provide a brief review of the susceptible-infectious-recovered (SIR) compartmental model.
In Section \ref{sec:model}, we develop a probabilistic state-space model for COVID-19 motivated by the classical SIR model.
In Section \ref{sec:inference}, we present strategies for posterior inference.
In Section \ref{sec:simu}, we carry out simulation studies to assess the performance of our method in estimating the epidemiological parameters. 
 In Section \ref{sec:case_studies}, we apply our method to COVID-19 data from six states of the United States (U.S.): Washington, New York, California, Florida, Texas, and Illinois.
We conclude with a discussion in Section \ref{sec:discussion}.

\section{Review of the Susceptible-Infectious-Recovered Model}
\label{sec:review_sir}

We start with a review of the susceptible-infectious-recovered (SIR) model \citep{kermack1927contribution, weiss2013sir}, a simple type of compartmental model.
The purpose of this review is to introduce the reader to the basics of epidemic modeling and motivate our proposed approach.

Consider a closed population of size $N$.
Here, ``closed'' means that $N$ does not vary over time. It is a good approximation for a fast-spreading and less fatal pandemic like COVID-19.
The SIR model divides the population into the following three compartments: 
\begin{enumerate}[noitemsep, nolistsep, leftmargin=8mm]
\item[(S)] Susceptible individuals: those who do not have the disease but may be infected;
\item[(I)] Infectious individuals: those who have the disease and are able to infect the susceptible individuals;
\item[(R)] Recovered/removed  individuals: those who had the disease but are then removed from the possibility of being infected again or spreading the disease.
Here, the removal can be due to several possible reasons, including death, recovery with immunity against reinfection, and quarantine and isolation from the rest of the population.
\end{enumerate}
At time $t$ ($t \geq 0$), denote by $S_t$, $I_t$ and $R_t$ the numbers of individuals in the S, I and R compartments, respectively, and write $\bV_t = (S_t, I_t, R_t)$. We have $S_t + I_t + R_t \equiv N$.

\subsection{Deterministic SIR Models}
The classical SIR model \citep{kermack1927contribution} describes the flow of people from S to I to R via the following system of differential equations:
\begin{align}
\frac{d S_t}{d t} = - \frac{\beta}{N} S_t I_t, \qquad 
\frac{d I_t}{d t} = \frac{\beta}{N} S_t  I_t - \alpha I_t, \qquad
\frac{d R_t}{d t} = \alpha I_t.
\label{eq:sir_model}
\end{align}
Here, $\beta$ is the \emph{disease transmission rate}, and $\alpha$ is the \emph{removal rate}.
The rationale behind the first equation in \eqref{eq:sir_model} is as follows: suppose each infectious individual makes \emph{effective contacts} (sufficient for disease transmission) with $\beta$ others per unit time; therefore, $\beta S /N$ of these contacts are with susceptible individuals per unit time, and as a result, $I$ infectious individuals lead to a rate of new infections $(\beta S  /N) \cdot I$.
The third equation in \eqref{eq:sir_model} describes that the infectious individuals leave the infective class at a rate of $\alpha I$.
The second equation in \eqref{eq:sir_model} follows immediately from the first and third equations.
The parameters $\beta$ and $\alpha$ are determined according to the natural history of the disease.
The quantities $\mathcal{R}_0 = \beta / \alpha$ and $\mathcal{R}_e = (\beta S_0) / (\alpha N) $ are referred to as the \emph{basic reproduction number} and \emph{effective reproduction number}, respectively, where $S_0$ is the initial number of susceptibles at time $t = 0$.

In some applications, it may be convenient to consider a  discrete-time approximation of the differential equations in Equation \eqref{eq:sir_model}, which can be expressed as follows:
\begin{align}
\begin{split}
S_t &= S_{t-1} - \beta S_{t-1} I_{t-1} / N, \\
I_t &= (1 - \alpha) I_{t-1} + \beta S_{t-1}  I_{t-1} / N, \\
R_t &= R_{t-1} + \alpha I_{t-1},
\end{split}
\label{eq:sir_model_discrete}
\end{align}
for $t = 1, 2, \ldots$. This discretization replaces the derivatives in Equation \eqref{eq:sir_model} by the differences per unit time.

The SIR models given by Equation \eqref{eq:sir_model} and \eqref{eq:sir_model_discrete} are both \emph{deterministic} models, meaning that their behaviors are completely determined by their initial conditions and parameter values.

\subsection{Stochastic SIR Models}

The deterministic SIR models are appealing due to their simplicity. 
However, the spread of disease is naturally stochastic. The disease transmission between two individuals is random rather than deterministic. Therefore, a stochastic formulation of the SIR model may be preferred for epidemic modeling, because it allows one to more readily capture the randomness of the epidemic process.

In a stochastic SIR model, $\{ \bV_t: t \geq 0 \}$ is treated as a stochastic process.
A commonly used formulation is as follows \citep{gibson1998estimating, o1999bayesian, andersson2000stochastic}.
Suppose that an infectious individual makes effective contacts with any given individual in the population at times given by a Poisson process of rate $\beta / N$, and assume all these Poisson processes are independent of each other.
Therefore, the expected number of effective contacts made by each infectious individual is $\beta$ per unit time.
Furthermore, suppose each infectious individual remains so (before being removed) for a period of time, known as the \emph{infectious period}.
Lastly, assume that the length of the infectious period for each individual is independent and follows an exponential distribution with mean $\alpha^{-1}$.
It can be shown that $\{ \bV_t: t \geq 0 \}$ is a Markov process with transition probabilities:
\begin{align}
\begin{split}
\text{Infection:} \quad  &\Pr[\bV_{t + \delta} = (s - 1, i + 1, r) \mid \bV_t = (s, i, r)] = \beta s i \delta / N + o(\delta), \\
\text{Removal:} \quad  &\Pr[\bV_{t + \delta} = (s, i - 1, r + 1) \mid \bV_t = (s, i, r)] = \alpha i \delta + o(\delta).
\end{split}
\label{eq:stochastic_sir}
\end{align}
Here, $\delta$ is a small increment in time.

\subsection{State-space SIR Models}
\label{sec:ss_sir}

There are, of course, other ways to model the uncertainty of the epidemic process.
Probabilistic state-space modeling approaches that build on deterministic models have recently been popular in the statistics literature \citep{dukic2012tracking, osthus2017forecasting, osthus2019dynamic}.
A state-space SIR model typically consists of two components: an evolution model for the epidemic process, and an observation model for the data.
As an example, the model in \cite{osthus2017forecasting} has the form
\begin{alignat*}{2}
\text{Evolution:}& \quad \bV_t  &&\sim p[ \bV_t \mid f(\bV_{t - 1}, \beta, \alpha), \kappa], \\
\text{Observation:}& \quad \tilde{I}_t  &&\sim p( \tilde{I}_t \mid I_t, \lambda),
\end{alignat*}
for $t = 1, 2, \ldots$.
In the evolution model, $f(\bV_{t - 1}, \beta, \alpha)$ is the solution to Equation \eqref{eq:sir_model} at time $t$ with a initial value of $\bV_{t - 1}$ at time $(t - 1)$ and parameters $\beta$ and $\alpha$, and $\bV_t$ is assumed to be centered at $f(\cdot)$ with its variance characterized by $\kappa$.
In other words, $\kappa$ measures the derivation of $\bV_{t}$ from the solution given by the deterministic model.
In the observation model, $\tilde{I}_t$ is the number of patients seen with the disease reported by healthcare providers, which can be thought of as a proxy to the true number of infectious individuals $I_t$. 
The observation $\tilde{I}_t$ is assumed to be centered at $I_t$ with variance characterized by $\lambda$.
State-space epidemic models are quite flexible and are in general more computationally manageable compared to stochastic epidemic models as in Equation \eqref{eq:stochastic_sir}.

The SIR model can be extended in many different ways, such as by considering vital dynamics (births and deaths) and demographics, adding more compartments to the model,  and allowing more possible transitions across compartments.
For example, the susceptible-exposed-infectious-recovered (SEIR) model includes an additional compartment for exposed individuals who are exposed to the disease but are not yet infectious, and the susceptible-infectious-recovered-infectious (SIRS) model allows recovered individuals to return to a susceptible state.
These extensions may better capture the characteristics of the disease under consideration.
For a comprehensive review of deterministic epidemic models, see, for example, \cite{anderson1991infectious}, \cite{hethcote2000mathematics} or \cite{Brauer2008}.
For a comprehensive review of stochastic epidemic models, see, for example, \cite{becker1999statistical}, \cite{andersson2000stochastic} or \cite{allen2008introduction}.

\section{Proposed Model for COVID-19}
\label{sec:model}

We now turn to our proposed model for the COVID-19 data, which belongs to the state-space model category (Section \ref{sec:ss_sir}).
Our approach integrates the discrete-time deterministic SIR model (Equation \ref{eq:sir_model_discrete}) and semiparametric Bayesian inference.
To capture some unique features of COVID-19, we consider the following extensions of the classical SIR model.
First, we split the infectious individuals into two subgroups: undocumented infectious individuals and documented infectious individuals.
The reason is that many people infected with SARS-CoV-2 have not been tested for the virus thus are not detected or reported as having the infection \citep{li2020substantial}.
Second, we allow some epidemiological parameters (such as the disease transmission rate $\beta$) to be time-varying to reflect the impact of mitigation policies such as stay-at-home orders and the change of public awareness of the disease over time.
We discuss details next.

\subsection{Model for the Epidemic Process}
\label{sec:process_model}

\begin{figure}[h!]
\centering
\includegraphics[width = 0.9\textwidth]{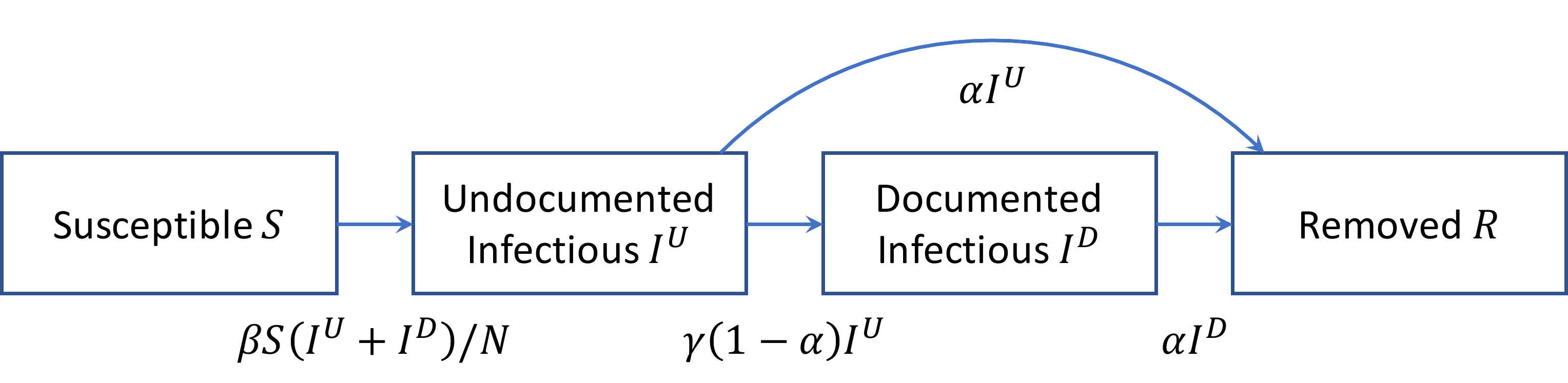} 
\caption{Compartmental model for COVID-19. We consider four compartments and four possible transitions across compartments. The number under each arrow indicates the transition rate between two compartments.}
\label{fig:proposed_model}
\end{figure}

Consider the transmission dynamics of COVID-19 in a specific country or region (e.g., a state, province or county). 
For simplicity, we consider a closed population (with no immigration and emigration) and also ignore nature births and deaths. Let $N$ denote the population size.
At any time point, we assume that each individual in the population precisely belongs to one of the following four compartments: 
\begin{enumerate}[noitemsep, nolistsep, leftmargin=10mm]
\item[(S)] Susceptible individuals who do not have the disease but are susceptible to it;
\item[(UI)] Undocumented infectious individuals who have the disease and may infect the susceptible individuals. However, they have not been detected as having the disease for several possible reasons. For example, they may have limited symptoms and are thus not tested for the disease;
\item[(DI)] Documented infectious individuals who have been confirmed as having the disease and are capable of infecting the susceptible individuals;
\item[(R)] Removed individuals who had the disease but are then removed from the possibility of being infected again or spreading the disease.
\end{enumerate}
We further assume that the infectious individuals (including both the UI- and DI-individuals) infect the S-individuals with a transmission rate of $\beta$. 
After being infected, a S-individual first becomes an UI-individual before being detected as a DI-individual.
All the infectious (UI- and DI-) individuals recover or die with a removal rate of $\alpha$. 
Those UI-individuals who have not been removed are diagnosed with the disease with a diagnosis rate of $\gamma$.
In total, there are four possible transitions across compartments: S to UI, UI to R, UI to DI, and DI to R. See Figure \ref{fig:proposed_model}.
Note that it is possible to assume different transmission rates for the UI- and DI-individuals, or to further split the UI and DI compartments into smaller subgroups (e.g., quarantined, hospitalized, etc.) with each subgroup having its distinct transmission rate. 
It is also possible to consider an extra compartment for the exposed (but not yet infectious) individuals as in the SEIR model.
Here, we use a more parsimonious model without the exposed compartment for simplicity and characterize the average transmission rate for all infectious individuals with a single parameter $\beta$ ($\beta$ depends on time, which will be clear later).
Finally, we assume recovery from COVID-19 confers immunity to reinfection, although there is only limited evidence for this assumption \citep{long2020antibody, kirkcaldy2020covid}.


We define day $t = 0$ as the date when the 100th case is confirmed in the country/region under consideration, and index subsequent dates by $t = 1, 2, \ldots, T$, where $T$ is the current date.
The reason for choosing day 0 in this way is because we believe the transmission dynamics of the disease is more trackable after a sufficient number of infectious individuals are reported in the country/region, although the choice of ``the 100th case'' is arbitrary and can be modified.
Denote by $S_{t}$, $I^U_{t}$, $I^D_{t}$ and $R_{t}$ the numbers of individuals belonging to compartments S, UI, DI and R on day $t$, respectively.
We have $S_{t} + I^U_{t} + I^D_{t} + R_{t} \equiv N$.
The transmission rate and diagnosis rate are allowed to vary over time and are hereafter denoted by $\beta_t$ and $\gamma_t$, respectively.
The number of individuals diagnosed with the disease between day $(t-1)$ and day $t$ is observed and is denoted by $B_{t-1}$.
This is our data.
We propose modeling the transmission dynamics of COVID-19 over time by the following equations:
\begin{align}
\begin{split}
S_{t} &= S_{t-1} -\beta_{t-1} S_{t-1} ( I^U_{t-1} +  I^D_{t-1} ) / N, \\
I^U_{t} &= (1 - \alpha) I^U_{t-1}  + \beta_{t-1} S_{t-1} ( I^U_{t-1} +  I^D_{t-1} ) / N - B_{t-1}, \\
I^D_{t} &= (1 - \alpha) I^D_{t-1} + B_{t-1}, \\
R_{t} &= R_{t-1} + \alpha (I^U_{t-1} + I^D_{t-1}),
\end{split}
\label{eq:covid_model}
\end{align}
for $t = 1, \ldots, T$.
Denote by $\bV_t = (S_t, I^U_{t}, I^D_{t}, R_t)$.
The epidemic process, $\{ \bV_t, t = 0, 1, \ldots, T \}$, is determined by its initial value $\bV_0$, the parameters $\{ \beta_t, \alpha \}$, and the observations $\{ B_t \}$.
Rigorously speaking, $\bV_t$ should be a vector of non-negative integers, but for computational convenience, we relax this restriction and only require it to be a vector of non-negative real numbers.
Model \eqref{eq:covid_model} is a simple extension of  \eqref{eq:sir_model_discrete} by adding a component of $I^U$, the undocumented infections, and by incorporating the observed daily new cases $B_{t-1}$ into the equations.
Later, we introduce a model for the observation $B_{t-1}$ to complete the state-space model.

With time-varying disease transmission rates, the basic reproduction number and effective reproduction number are also functions of time. That is, $\mathcal{R}_0(t) = \beta_t / \alpha$ and 
\begin{align*}
\mathcal{R}_e(t) = (\beta_t S_t) / (\alpha N).
\end{align*}
Here, $\mathcal{R}_e(t)$ is interpreted as the
rate of secondary infections generated by each infectious case at time $t$, scaled by the length of the infectious period ($\alpha^{-1}$).
If $\mathcal{R}_e(t) < 1$ for $t \geq t^*$, then the number of infectious individuals $(I^U_t + I^D_t)$ will monotonically decrease after time $t^*$, because each infectious individual will only be able to infect less than 1 other during the course of his/her infectious period.
In other words, an $\mathcal{R}_e(t) < 1$ indicates containment of the disease.
Due to the important role of $\mathcal{R}_e(t)$ in characterizing disease spread, we consider the estimation of $\mathcal{R}_e(t)$ as our main interest.

\subsection{Model for the Observed Data}

Our observations only consist of the daily new confirmed COVID-19 cases, $B_t$.
Assume that on day $t$, the UI-individuals who have not been removed are diagnosed with the disease with a diagnosis rate of $\gamma_t$. Mathematically, this means
$B_{t} = \gamma_t (1 - \alpha) I^U_{t}$,
where $\gamma_t$ is between 0 and 1. We consider the logit transformation of $\gamma_t$,  $\tgamma_t = \logit(\gamma_t) \triangleq \log[ \gamma_t / (1 - \gamma_t)]$.  Other transformations, such as the probit and complementary log-log transformations, can also be specified in the \texttt{BaySIR} package. Empirically we find the proposed model to be robust to different specifications of the link function (See Appendix \ref{supp:sec:sens_analysis}). We assume a prior transformation,
\begin{align}
\tgamma_t \sim \N (\by_t^{\top} \bmeta, \sigma_{\gamma}^2),
\label{eq:diag_rate}
\end{align}
where $\by_{t}$ is a vector of covariates that are thought to be related to the diagnosis rate. 
In other words, the sampling model for $B_t$ can be written as
\begin{align}
\logit \left[ \frac{B_t}{(1 - \alpha) I^U_{t}} \right]  \mid I^U_{t}, \alpha \sim \N (\by_t^{\top} \bmeta, \sigma_{\gamma}^2).
\label{eq:logit_link}
\end{align} 
In the simulation studies and real data analyses, we use a simple choice of $y_{t} = 1$, assuming the mean diagnosis rate is a constant.
It is possible to include other covariates in $\by_{t}$, such as the number of tests, but empirically we find it hard to detect the effects of these covariates.
In the \texttt{BaySIR} package, the user has the option to include any covariates.
The parameters $\bmeta$ and $\sigma_{\gamma}^2$ are the regression coefficients and variance term, respectively,
where $\sigma_{\gamma}^2$ captures random fluctuations of confirmed case counts and report errors.


For some countries and regions, the numbers of recoveries and deaths are also available, and one may think of using them as the observed number of removed individuals.
We choose not to use these data for two reasons.
First, many infected individuals, even with confirmed disease, are not hospitalized, and their recoveries are not recorded.
In other words, the reported number of recoveries and deaths is a significant underestimate of the size of the removed population.
Second, according to \cite{wolfel2020virological}  and \cite{he2020temporal}, the ability of a COVID-19 patient to infect others becomes negligible several days before the patient recovers or dies, suggesting that ``removal'' in our application is not equivalent to ``recovery or death''.

\subsection{Prior Specification}

In what follows, we discuss prior specification for the initial condition and parameters. 
Due to the limited amount of observable information,  many latent variables and parameters in the proposed model are unidentifiable.
See Appendix \ref{supp:sec:identifiability} for a detailed discussion with an example showing that two epidemic processes with distinct parameters lead to exactly the same observed data.
We note that this problem is pervasive in most existing methods, and a typical solution to the problem is to prespecify some parameter values based on prior knowledge.
Here, we elicit informative priors for some parameters based on the clinical characteristics of COVID-19, which favor more clinically plausible estimates.

\paragraph{Initial condition}
The initial condition of the epidemic process refers to the vector $\bV_0 = (S_0, I^U_{0}, I^D_{0}, R_0)$.
We assume that there are no removed individuals on day $0$, i.e., $R_{0} = 0$. 
As a result, the number of DI-individuals on day $0$, $I_{0}^D$, equals to the cumulative number of confirmed cases on that day and is observed.
We further assume
\begin{align*}
I_{0}^U / I_{0}^D \sim \Ga(\nu_1, \nu_2),
\end{align*}
where $\Ga(\nu_1, \nu_2)$ refers to a gamma distribution with shape and rate parameters $\nu_1$ and $\nu_2$, respectively.
We set $\nu_1 = 5$ and $\nu_2 = 1$, such that $\E(I_{0}^U / I_{0}^D) = 5$. This choice is based on the findings in \cite{li2020substantial} that 86\% of all infections were undocumented at the beginning of the epidemic in China.
Lastly, note that $S_0 = N - I_{0}^U - I_{0}^D - R_0$.

\paragraph{Transmission rate}
The disease transmission rate $\beta_{t}$ must be non-negative. We consider $\tbeta(t) = \log(\beta_{t})$ and assume
\begin{align*}
\tbeta(t) \sim \GP[m(t), C(t, t')],
\end{align*}
where $\GP[m(t), C(t, t')]$ refers to a Gaussian process (GP) with mean function $m(t)$ and covariance function $C(t, t')$.
The GP \citep{rasmussen2006gaussian} is a very flexible prior model for a stochastic process. It enables one to capture potential non-linear relationships between $t$ and $\tbeta(t)$ without the need to impose any parametric assumptions. Specifically, for any $t_1, \ldots, t_n \geq 0$, the vector $(\tbeta(t_1), \ldots, \tbeta(t_n))^{\top}$ follows a multivariate Gaussian distribution with mean $(m(t_1), \ldots, m(t_n))^{\top}$ and covariance matrix $\mathbf{C}$ with the $(i,j)$-th entry being $C(t_i, t_j)$. For applications of GP to epidemic modeling, see, for example, \cite{xu2016bayesian} and \cite{kypraios2018bayesian}.

We specify $m(t)$ and $C(t, t')$ as below:
\begin{align}
m(t) = \bx_t^{\top} \bmu, \qquad
C(t, t') = \sigma_{\beta}^2 \rho^{| t - t' |}.
\label{eq:GP_specification}
\end{align}
Here, $\bx_{t}$ is a vector of covariates that are thought to be related to the transmission rate, and $\bmu$ is a vector of regression coefficients. 
In the simulation studies and real data analyses, we use $\bx_{t} = (1, t)^{\top}$, which contains an intercept term and the time. Other covariates, such as indicators for  mitigation policies at time $t$, may also be included in $\bx_{t}$.
Nevertheless, in practice, we find our GP model with a time trend is sufficient to capture the change of $\tbeta(t)$ over time and the potential effects of mitigation policies and public awareness.
Users of our software may include other covariates using the R package \texttt{BaySIR}.
The variance parameter $\sigma_{\beta}^2$ characterizes the amplitude of the difference between $\tbeta(t)$ and $m(t)$, and the correlation parameter $\rho$ characterizes the correlation between $\tbeta(t)$ and $\tbeta(t')$ for any $t$ and $t'$.
We note that based on our specification of the covariance function, our GP model is equivalent to a first-order autoregressive model. Indeed, autoregressive models of any orders are discrete-time equivalents of GP models with Mat\'{e}rn covariance functions \citep{roberts2013gaussian}.

We place the following priors on $\bmu$, $\sigma_{\beta}$ and $\rho$:
\begin{align*}
\bmu \sim \N(\bmu^*, \Sigma_{\mu}), \quad
\sigma_{\beta}^2 \sim \IG(11, 1), \quad
\rho \sim \Be(4, 1),
\end{align*}
such that $\E(\sigma_{\beta}^{2}) = 0.1$ and $\E(\rho) = 0.8$. Here, $\IG( \cdot, \cdot )$ refers to an inverse gamma distribution, and $\Be( \cdot, \cdot )$ refers to a beta distribution. The prior choices for $\sigma_{\beta}^2$ and $\rho$ shrink $\tbeta(t)$ toward its mean function (i.e., a linear regression model) and impose a strong prior correlation between the transmission rates for two consecutive days.
For the prior of $\bmu$, we use $\bmu^* = (-1.31, 0)^{\top}$ and $\Sigma_{\mu} = \text{diag}(0.3^2, 1^2)$, where $\text{diag}( \cdot )$ represents a diagonal matrix.
In this way, the prior median of the basic reproduction number on day 0 is 2.5 (with 95\% credible interval 1.4 to 4.5), assuming the infectious period is 9.3 days. This is based on the findings in \cite{li2020early} and \cite{wu2020nowcasting}.
The prior also induces a mild shrinkage (towards 0) for the regression coefficient of the time trend.

\paragraph{Removal rate} 
The removal rate is between 0 and 1. The inverse of the removal rate, $\alpha^{-1}$, corresponds to the average time to removal after infection. We assume
\begin{align*}
\alpha^{-1} \sim \Ga(\nu^{\alpha}_1, \nu^{\alpha}_2) \cdot \mathbbm{1}(\alpha^{-1} \geq 1).
\end{align*}
We take $\nu^{\alpha}_1 = 325.5$ and $\nu^{\alpha}_1 = 35$, such that $\E(\alpha^{-1}) = 9.3$ with prior 95\% credible interval between $8.3$ and $10.3$ days.
The mean infectious period of 9.3 days is chosen based on the findings in \cite{he2020temporal}, who estimated that the infectiousness of COVID-19 starts from around 2.3 days before symptom onset and declines quickly within 7 days after symptom onset.

\paragraph{Diagnosis rate}
We place the following standard weakly informative priors on $\bmeta$ and $\sigma_{\gamma}^2$, the regression coefficients and variance term in the diagnosis rate model (Equation \ref{eq:diag_rate}):
\begin{align*}
\bmeta \sim \N(\bmeta^*, \Sigma_{\eta}), \quad
\sigma_{\gamma}^2 \sim \IG(1, 1).
\end{align*}
When $y_t$ only has an intercept term, we use $\eta \sim \N(0, 1^2)$.

\section{Inference}
\label{sec:inference}

\subsection{Posterior Sampling}
Let $\btheta = \{ I^U_0, \bbeta, \alpha, \bmu, \sigma_{\beta}, \rho, \bmeta, \sigma_{\gamma}^2 \}$ denote all model parameters and hyperparameters, where $\bbeta = (\beta_0, \beta_1, \ldots, \beta_T)$, and let $\bB = (B_0, B_1, \ldots, B_T)$ be the vector of daily increments in confirmed cases.
The joint posterior distribution of $\btheta$ is given by
\begin{align*}
\pi(\btheta \mid \bB, I^D_0) \propto \left[ \prod_{t = 0}^T \phi(\tgamma_t \mid \by_t^{\top} \bmeta, \sigma_{\gamma}^2) \right] \cdot \pi^* (\btheta)
\end{align*}
where $\phi( \cdot \mid \mu, \sigma^2)$ denotes the density function of a normal distribution with mean $\mu$ and standard deviation $\sigma^2$, and $\pi^*(\btheta)$ represents the prior density of $\btheta$.
Recall that $\tgamma_t = \logit \left\{ B_t / [ (1 - \alpha) I^U_t] \right\} $.

We use a Markov chain Monte Carlo (MCMC) algorithm (see, e.g., \citealp{liu2008monte}), in particular the Gibbs sampler, to simulate from the posterior distribution and implement posterior inference.
Metropolis-Hastings steps are used when the conditional posterior distribution of a parameter is not available in closed form.
The regular Gibbs sampler is not very efficient in our application because of the strong correlations among the model parameters. 
This issue was also noted by \cite{osthus2017forecasting}.
We therefore use parallel tempering (PT)  to  improve the convergence and mixing of the Markov chains \citep{geyer1991markov}.
Consider $J$ parallel Markov chains with a target distribution of
\begin{align*}
\pi_j (\btheta_j \mid \bB, I^D_0) \propto \left[ \prod_{t = 0}^T \phi(\tgamma_{t, j} \mid \by_t^{\top} \bmeta_j, \sigma_{\gamma, j}^2) \right]^{1 / \Delta_j} \cdot \pi^*(\btheta_j)
\end{align*}
for the $j$-th chain, where $\Delta_j$ is the \emph{temperature}.
The temperatures $\{ \Delta_1, \Delta_2, \ldots, \Delta_J \}$ are decreasing with $\Delta_J = 1$.
Thus the target distribution of the $J$-th chain is the original posterior $\pi(\btheta \mid \bB, I^D_0)$.
At each MCMC iteration, we first independently update all $J$ chains based on Gibbs transition probabilities.
Then, for $j = 1, 2, \ldots, J - 1$, we propose a swap between $\btheta_j$ and $\btheta_{j+1}$ and accept the proposal with probability
\begin{align*}
p_{\text{swap}}(\btheta_j, \btheta_{j+1}) = 1 \wedge \left[ \prod_{t = 0}^T \frac{\phi(\tgamma_{t, j+1} \mid \by_t^{\top} \bmeta_{j+1}, \sigma_{\gamma, j+1}^2) }{\phi(\tgamma_{t, j} \mid \by_t^{\top} \bmeta_j, \sigma_{\gamma, j}^2) } \right]^{\frac{1}{\Delta_j} - \frac{1}{\Delta_{j+1}}}.
\end{align*}
The draws from the $J$-th chain are kept.
A chain with a higher temperature can more freely explore the posterior space, and the swap proposal allows interchange of states between adjacent chains.
Therefore, the PT scheme helps the Markov chain avoid getting stuck at local optima.
In Appendix \ref{supp:sec:ptmcmc}, we demonstrate the advantage of the PT scheme with an example.

In the simulation studies and real data analyses, we run $J = 10$ parallel Markov chains with a temperature of $\Delta_j = 1.5^{10-j}$ for the $j$-th chain.
We run MCMC simulation for 50,000 iterations, discard the first 20,000 draws as initial burn-in, and keep one sample every 30 iterations. This leaves us a total of 1,000 posterior samples.

\subsection{Predictive Inference}
\label{sec:predictive}

In addition to the estimation of epidemiological parameters, one may be interested in the prediction of a future observation, which can be achieved by sampling from its posterior predictive distribution.
As an example, let $\bB^* = (B_{T+1}, \ldots, B_{T + T^*})$ denote the vector of daily confirmed cases for future days $t = T + 1, \ldots, T + T^*$.
The posterior predictive distribution of $\bB^*$ is given by
\begin{align}
\pi( \bB^* \mid \bB, I^D_0) = \int \pi( \bB^* \mid \btheta, \bB, I^D_0) \cdot \pi(\btheta \mid \bB, I^D_0) \, \text{d}\btheta.
\label{eq:post_pred}
\end{align}
Sampling from \eqref{eq:post_pred} involves computing $\pi( \tilde{\bbeta}^* \mid \bB, I^D_0) = \int \pi( \tilde{\bbeta}^* \mid \tilde{\bbeta}) \cdot \pi( \tilde{\bbeta} \mid \bB, I^D_0) \, \text{d}\tilde{\bbeta}$ for
$\tilde{\bbeta}^* = (\tbeta_{T+1}, \ldots, \tbeta_{T + T^*})$. We have 
\begin{align*}
\tilde{\bbeta}^* \mid \tilde{\bbeta} \sim \N \left[ \mathbf{X}^* \bmu + \mathbf{C}^* \mathbf{C}^{-1} (\tilde{\bbeta} - \mathbf{X} \bmu),  \mathbf{C}^{**} - \mathbf{C}^* \mathbf{C}^{-1} \mathbf{C}^* \right],
\end{align*}
where $\mathbf{X} = (\bx_0, \ldots, \bx_T)^{\top}$, $\mathbf{X}^* = (\bx_{T+1}, \ldots, \bx_{T + T^*})^{\top}$, $\mathbf{C}^*$ is a $T^* \times (T + 1)$ matrix with the $(i, j)$-th entry being $C(T+i, j-1)$, and  $\mathbf{C}^{**}$ is a $T^* \times T^*$ matrix with the $(i, j)$-th entry being $C(T+i, T+j)$.
This is based on a GP prediction rule \citep{rasmussen2006gaussian}.

\section{Simulation Studies}
\label{sec:simu}

\begin{figure}[h!]
\centering
\begin{tabular}{ccc}
\includegraphics[width = 0.31\textwidth]{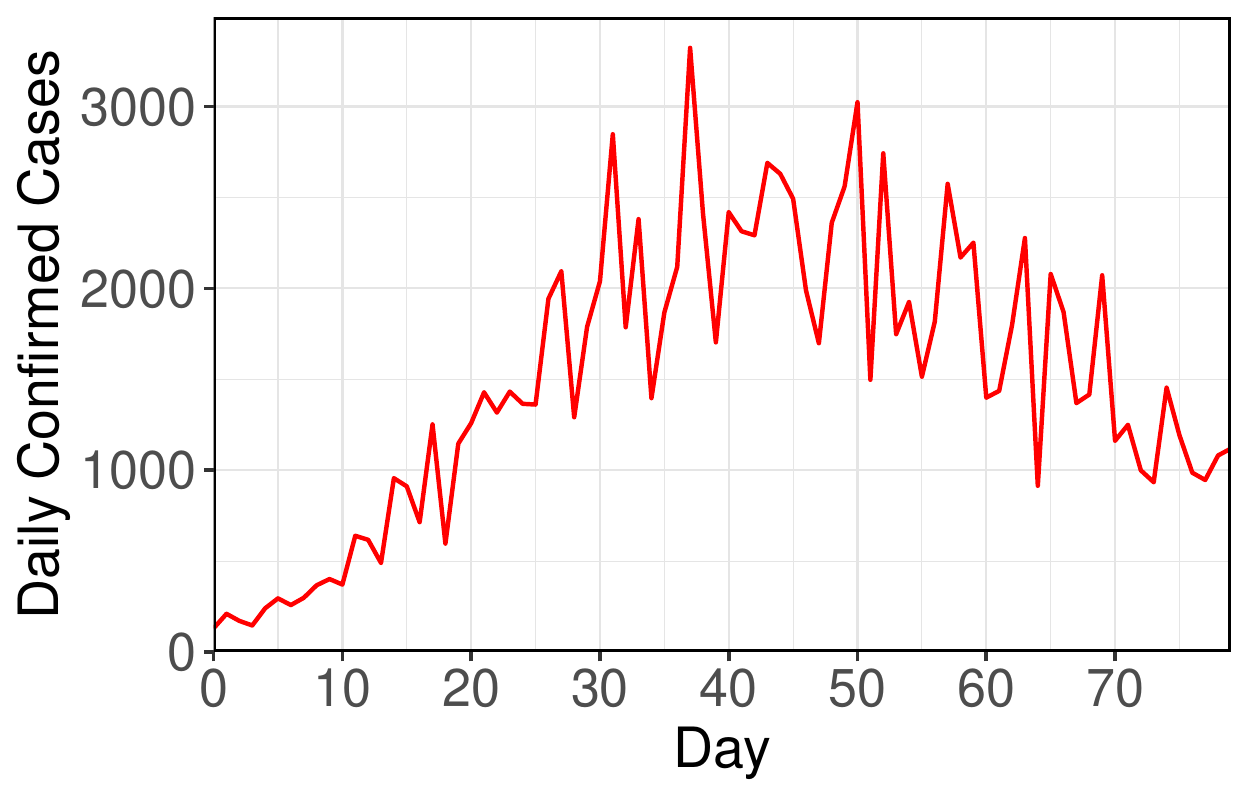} &
\includegraphics[width = 0.31\textwidth]{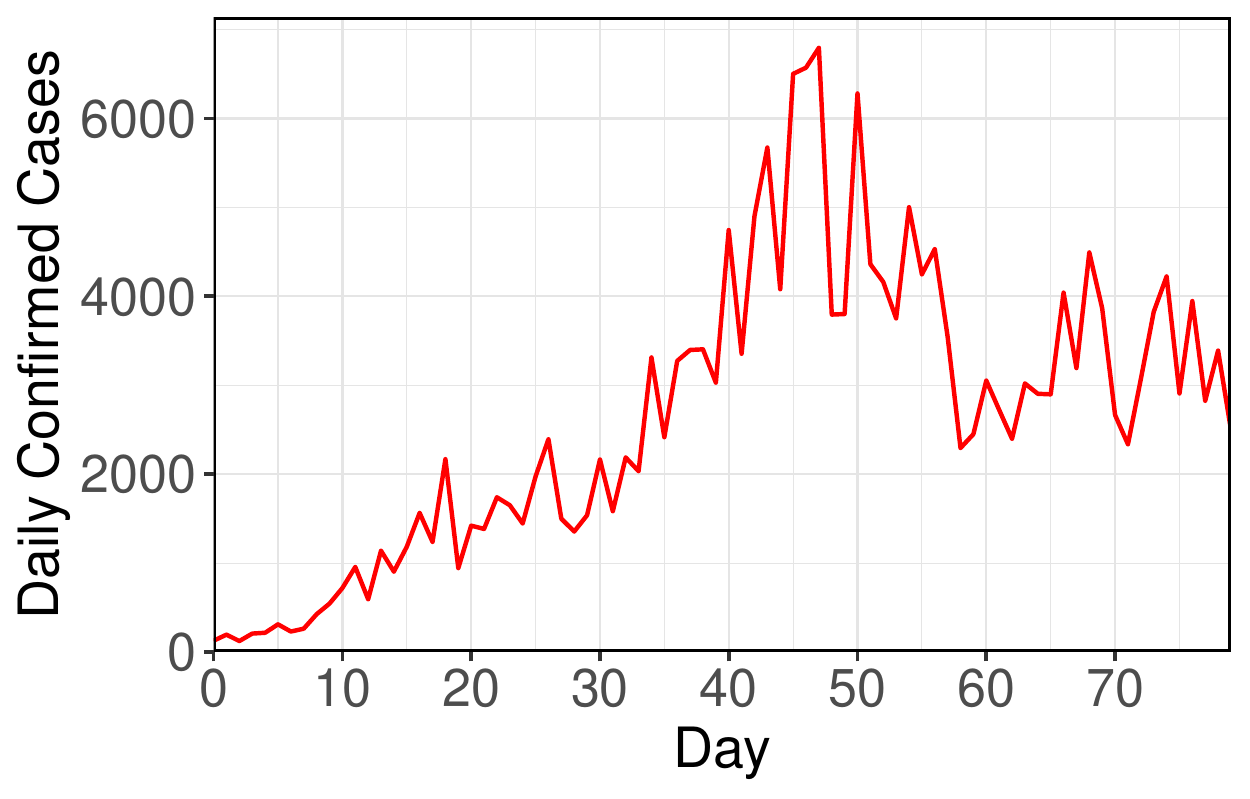} &
\includegraphics[width = 0.31\textwidth]{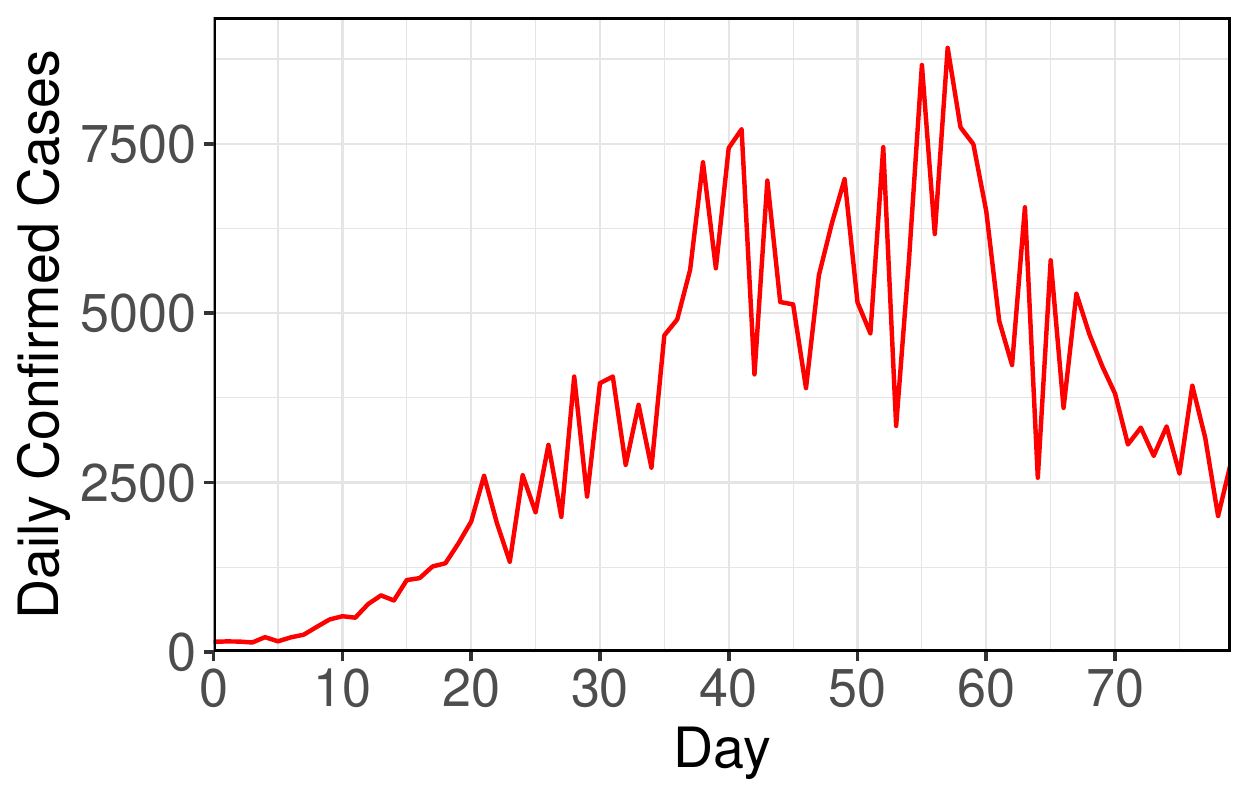} \\
\includegraphics[width = 0.31\textwidth]{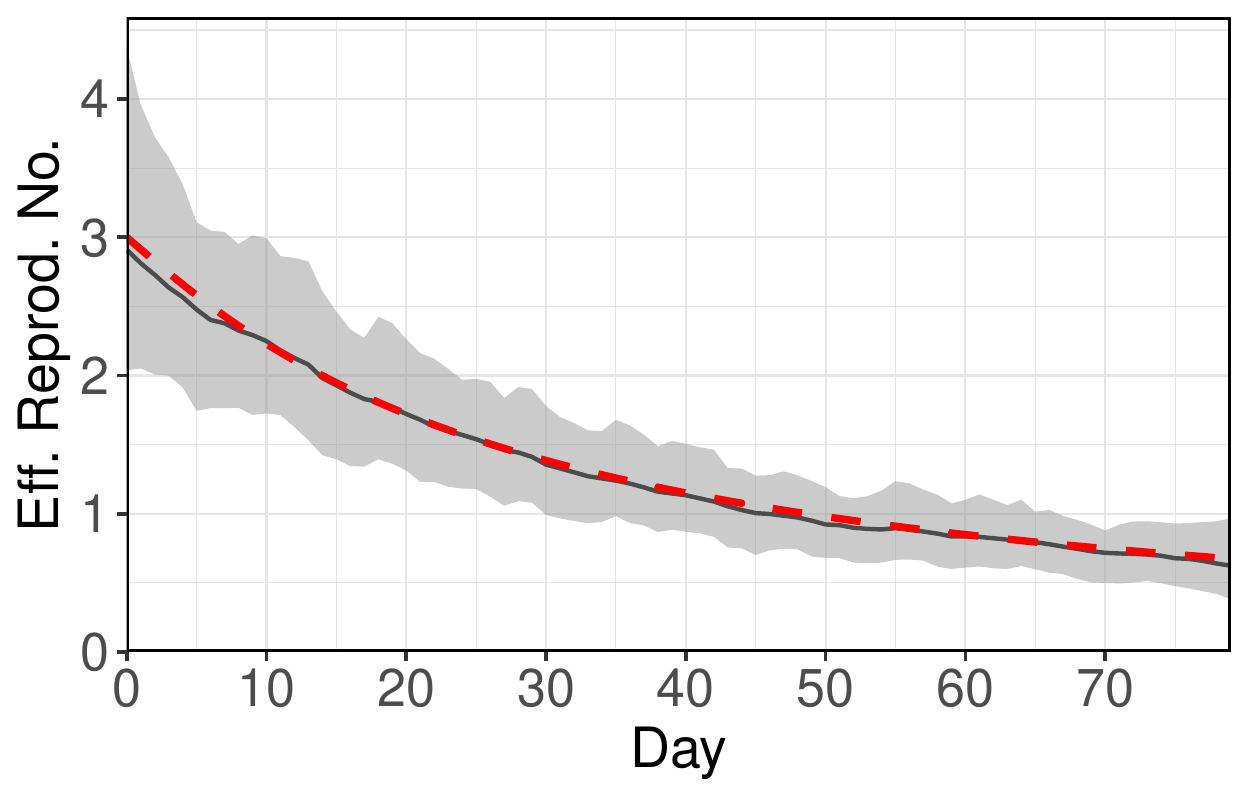} &
\includegraphics[width = 0.31\textwidth]{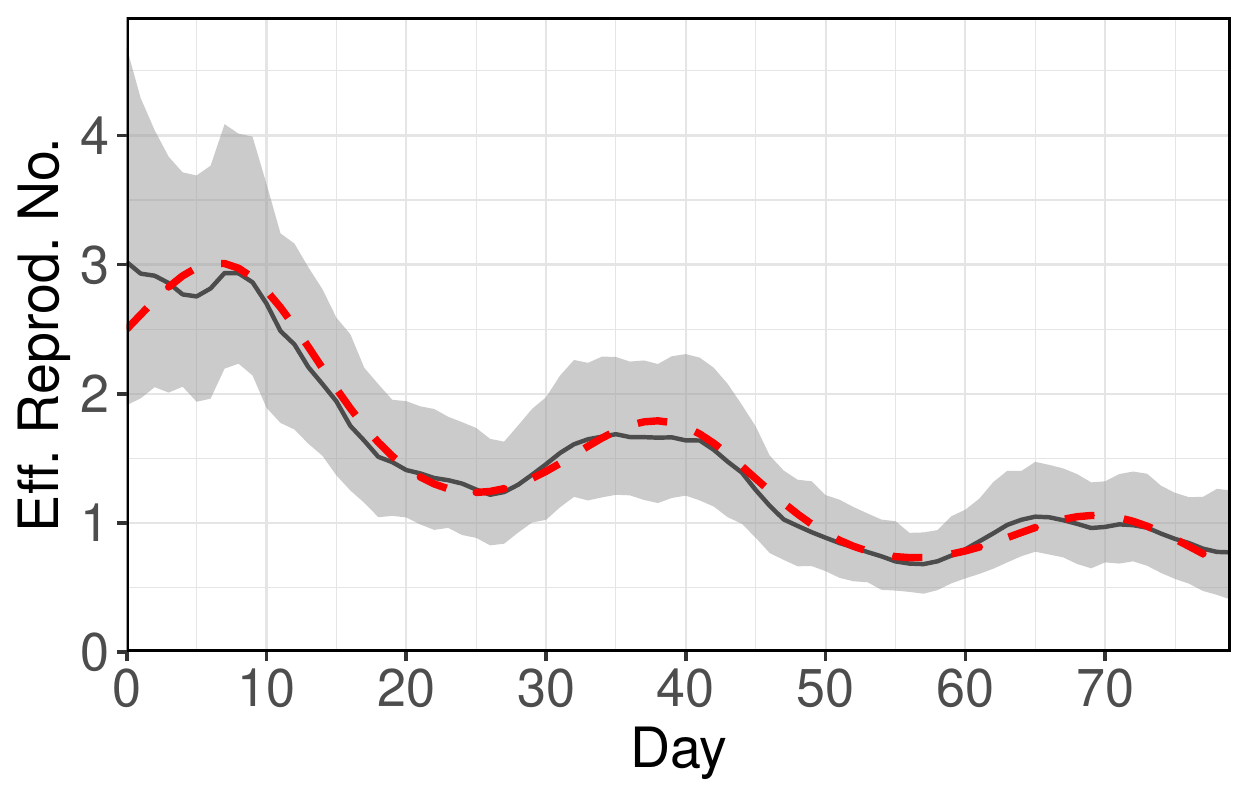} &
\includegraphics[width = 0.31\textwidth]{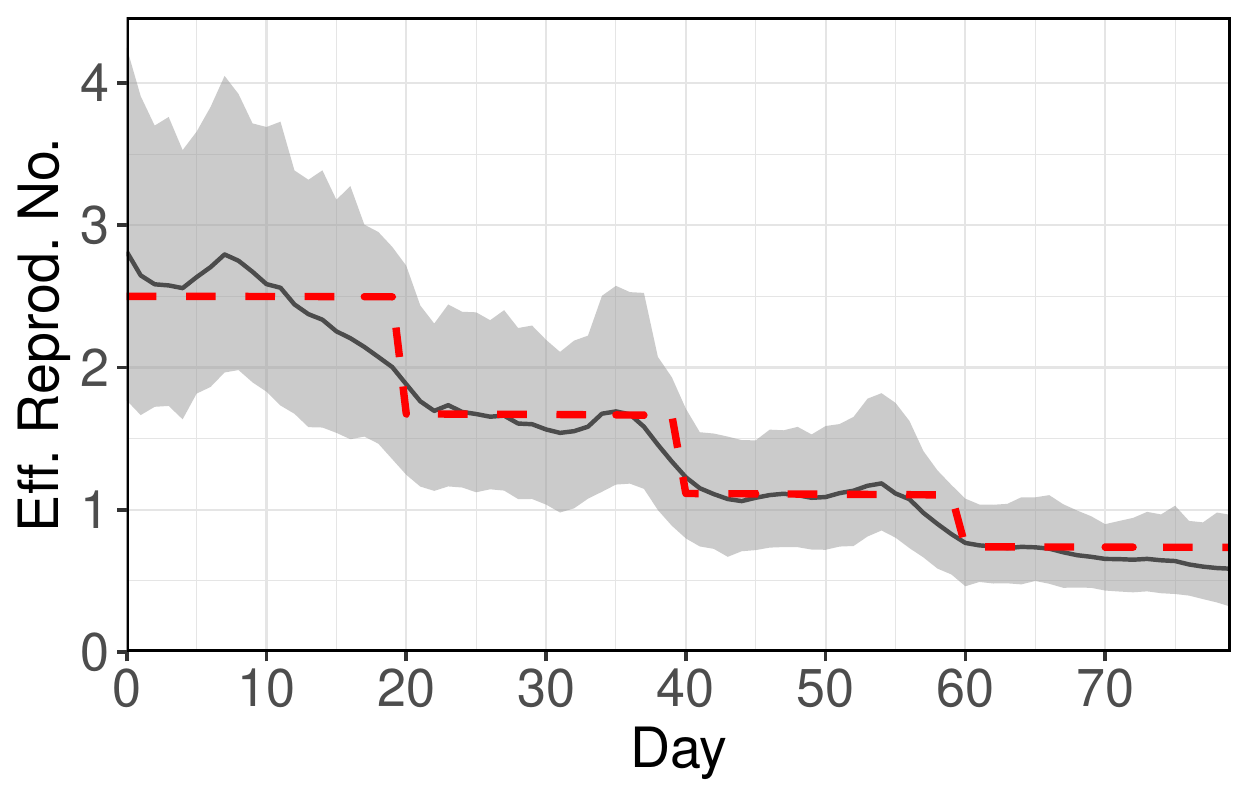} \\
(a) Scenario 1 & 
(b) Scenario 2 &
(c) Scenario 3
\end{tabular}
\caption{The upper panel shows the simulated daily confirmed cases for the three scenarios. 
The lower panel shows the estimated time-varying effective reproduction numbers (solid black line), 95\% credible intervals (grey band), and simulation truth (dashed red line) for the three scenarios.}
\label{fig:simu}
\end{figure}

We assess the performance of the proposed method in estimating the epidemiological parameters by applying it to simulated epidemic time series.
Consider a closed population of size $N = 20,000,000$. 
We assume the initial condition on day $0$ is $I^D_0 = 100$, $I^U_0 = 800$, $R_0 = 0$, and $S_0 = N - I^D_0 - I^U_0$.
We set the removal rate $\alpha = 9.3^{-1}$.
For the transmission rate, we consider the following three scenarios:
\begin{enumerate}[noitemsep, nolistsep, leftmargin=17mm]
\item[(Scn. 1)]  $\beta_t = b \cdot \alpha / [(t + 1)^c - a]$, where $a, b$ and $c$ are chosen such that $\mathcal{R}_0(0) = 3$, $\mathcal{R}_0(14) = 2$ and $\mathcal{R}_0(49) = 1$;
\item[(Scn. 2)]  $\beta_t = \alpha \cdot \exp \left[ a \cdot \sin(0.2 t) - bt + c \right]$, where $a, b$ and $c$ are chosen such that $\mathcal{R}_0(0) = 2.5$, $\mathcal{R}_0(14) = 2.2$ and $\mathcal{R}_0(49) = 1$;
\item[(Scn. 3)]  $\beta_t = \alpha \cdot \exp \big[ \log(2.5) -  0.4 \cdot \lfloor (t/20) \rfloor \big]$, where $\lfloor a \rfloor$ represents the largest integer that is smaller than $a$.
\end{enumerate}
Recall that $\mathcal{R}_0(t) = \beta_t / \alpha$.
In all the scenarios, $\mathcal{R}_0(t) \rightarrow 0^+$ as $t \rightarrow \infty$. For scenario 2, $\mathcal{R}_0(t)$ is non-monotonic, and for scenario 3, $\mathcal{R}_0(t)$ is discontinuous.
Next, we generate $\tgamma_t \sim \N \left[ \logit(0.2), 0.25^2 \right]$ and $\gamma_t = 1 - \exp(-\exp(\tgamma_t))$.
Finally, for each scenario, we generate a hypothetical epidemic process for 80 days according to Equation \eqref{eq:covid_model} with $B_t =  \gamma_t (1 - \alpha) I^U_{t}$.
We keep $\bB = (B_0, \ldots, B_T)$ and $I^D_0$ as our observations ($T = 79$).
The simulated datasets, shown in Figure \ref{fig:simu} (upper panel), are similar to a real COVID-19 dataset (e.g., Figure \ref{fig:within_forecast}).

We fit the proposed model to the simulated datasets using the PTMCMC algorithm. 
Figure \ref{fig:simu} (lower panel) shows a comparison of the estimated time-varying effective reproduction numbers with the simulation truth.
The simulation truth is nicely recovered, and the 95\% credible intervals of $\mathcal{R}_e(t)$'s always cover the true values.

We also carry out sensitivity analyses to explore how the choice of the link function (Equation \ref{eq:logit_link}) and priors can affect the performance of the proposed method. Details of the sensitivity analyses are reported in Appendix \ref{supp:sec:sens_analysis}. 
In general, our method is robust to different specifications of the link function.
The choice of the priors, on the other hand, may have an impact on the parameter estimates, because of parameter unidentifiability issues (see Appendix \ref{supp:sec:identifiability}).

\section{Case Studies}
\label{sec:case_studies}

To illustrate the practical application of the proposed method, we carry out data analysis based on daily counts of confirmed COVID-19 cases reported by JHU CSSE. 
This is the $B_t$ in our model.
We limit our analysis to six U.S. states (Washington, New York, California, Florida, Texas, and Illinois) to keep the paper in reasonable length.
The reader can carry out independent analysis for other states, countries or regions using the R package \texttt{BaySIR}.
The populations of these states are obtained from U.S. Census Bureau.

\subsection{Estimation of the Effective Reproduction Number} 

\begin{figure}[h!]
\centering
\includegraphics[width = 0.31\textwidth]{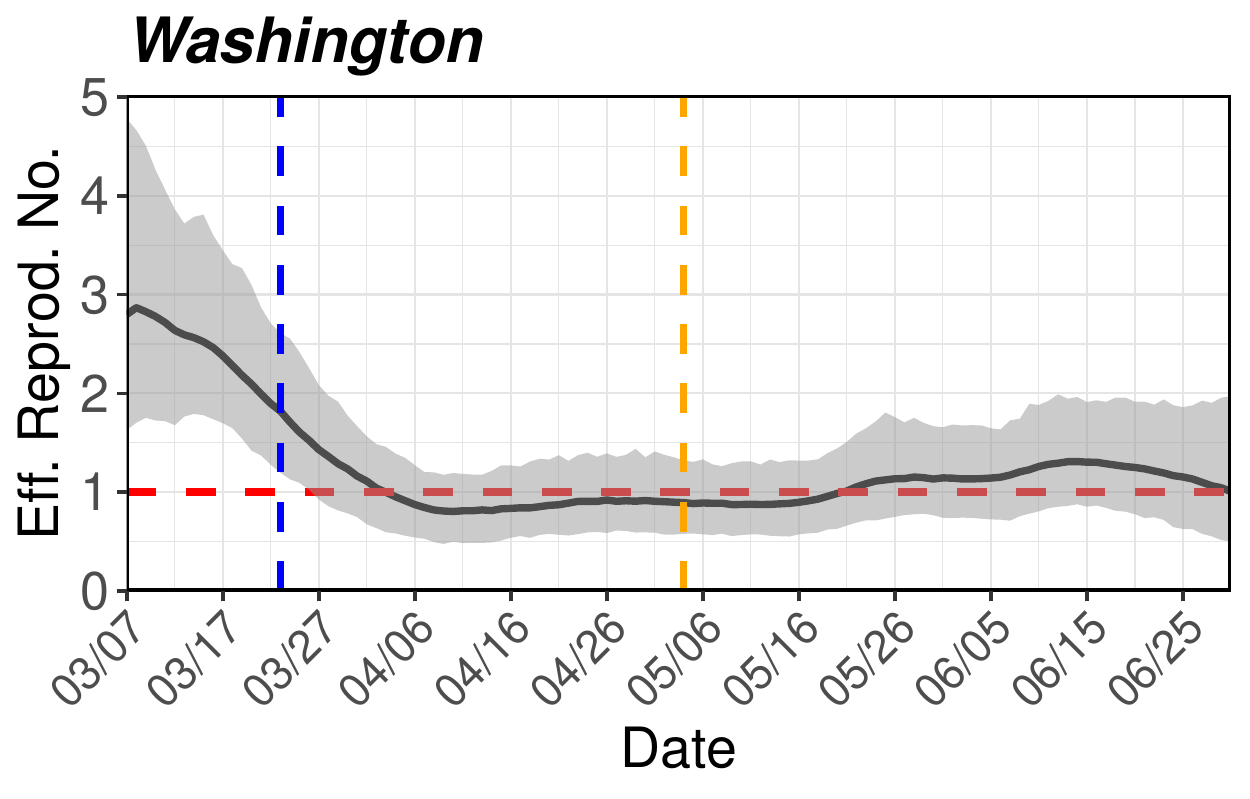} 
\includegraphics[width = 0.31\textwidth]{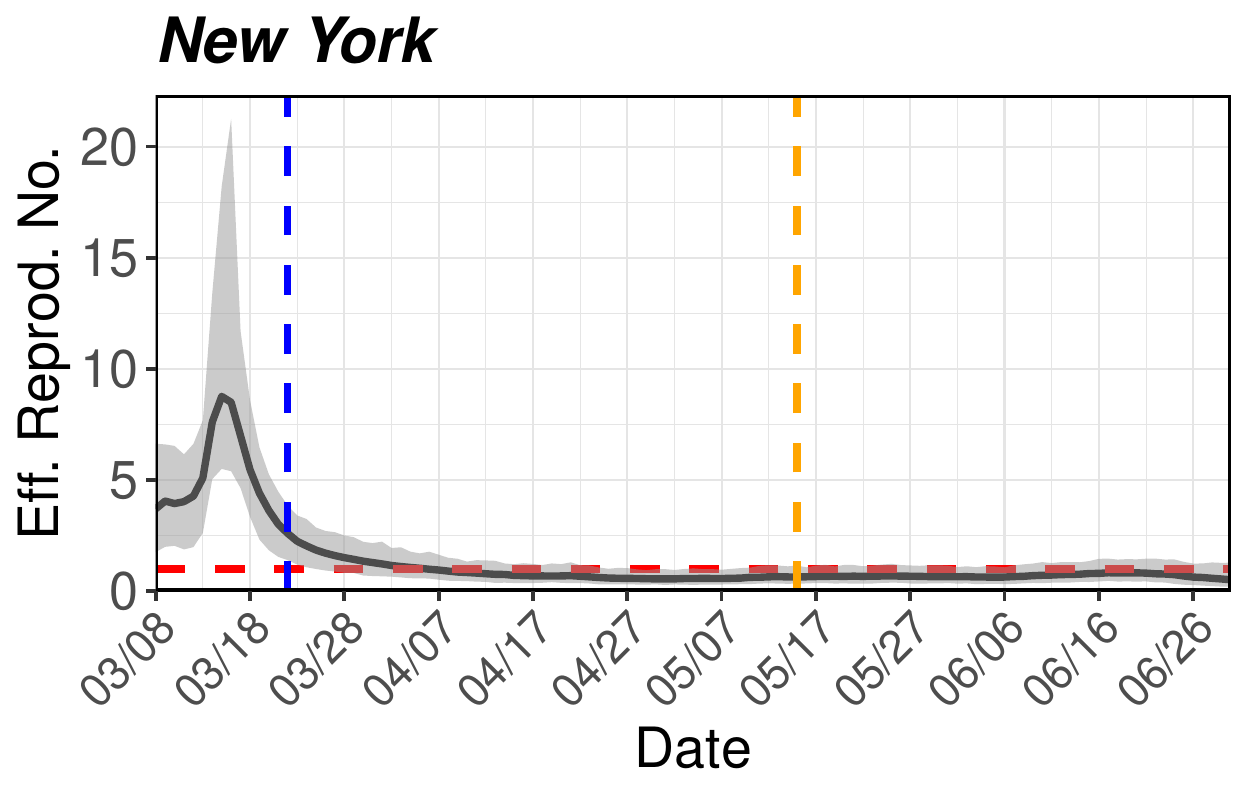} 
\includegraphics[width = 0.31\textwidth]{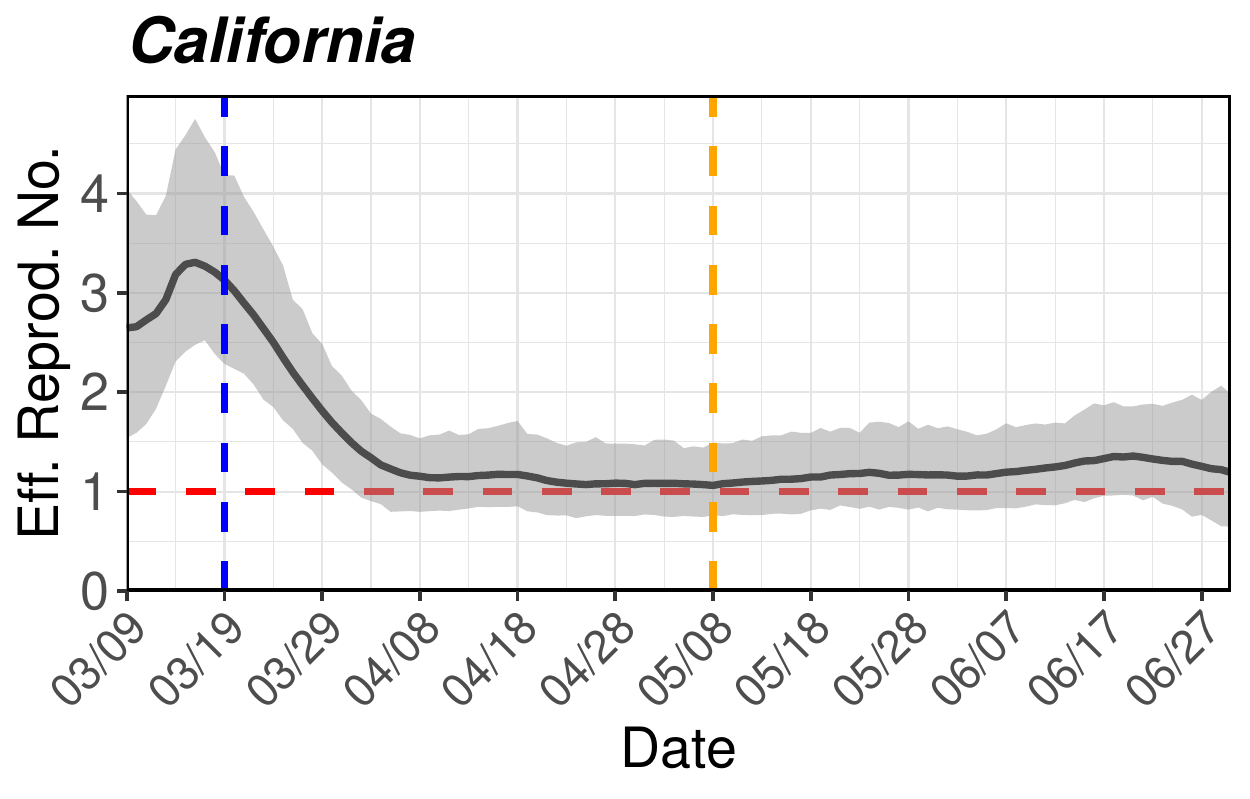} 
\includegraphics[width = 0.31\textwidth]{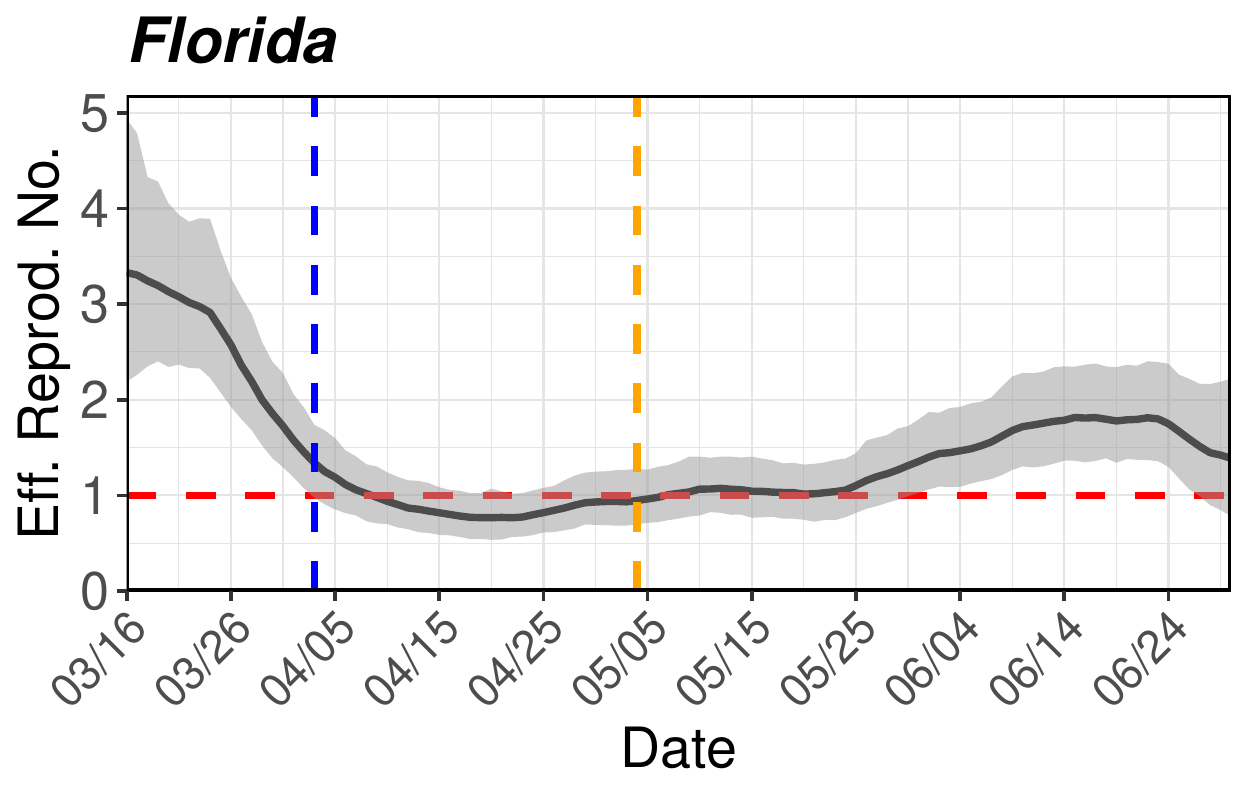}
\includegraphics[width = 0.31\textwidth]{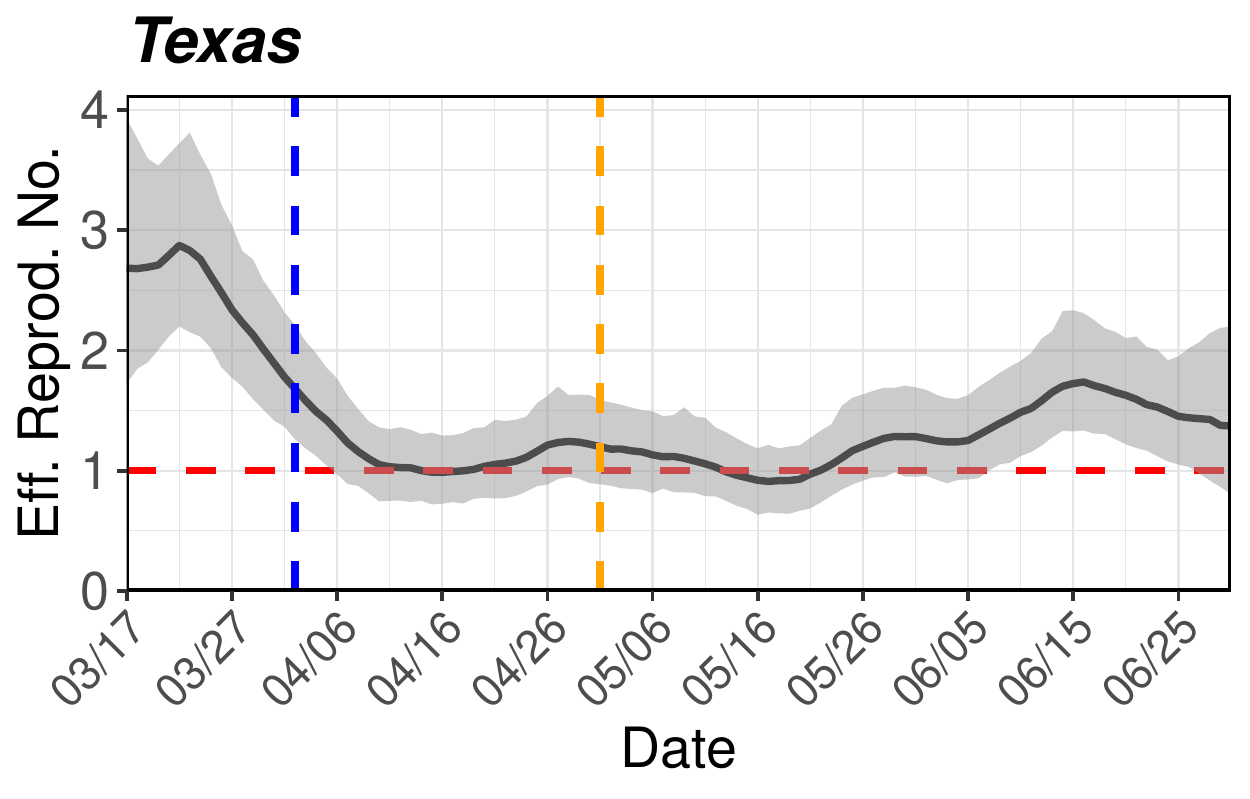}
\includegraphics[width = 0.31\textwidth]{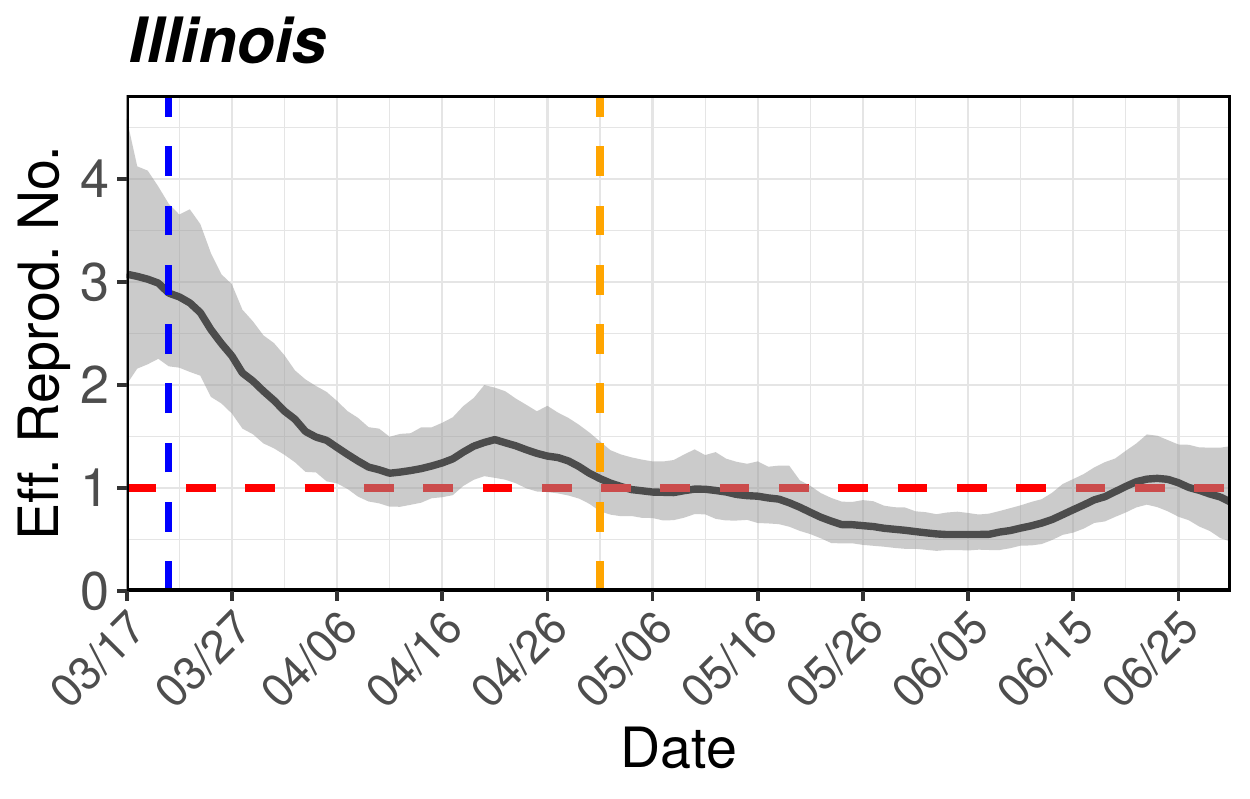} 
\caption{Estimated time-varying effective reproduction numbers (solid black line) for six U.S. states: Washington, New York, California, Florida, Texas, and Illinois. The start date in each graph is the date when the 100th case is confirmed in the state.
The grey band represents the 95\% posterior credible interval.
The dashed vertical lines correspond to the start dates of statewide stay-at-home orders and state reopening plans. The dashed horizontal line represents an $\mathcal{R}_e$ of 1. }
\label{fig:R_eff}
\end{figure}

Figure \ref{fig:R_eff} shows the estimated $\mathcal{R}_e(t)$ for the six states. 
The start dates of statewide stay-at-home orders and state reopening plans are also displayed in the figure for reference (data source: \citealp{mervosh2020see} and \citealp{washington2020where}).
The estimated initial $\mathcal{R}_e$ ranges from 2.5 to 4.0.
Specifically, $\mathcal{R}_e(0) = 2.8$, $3.7$, $2.6$, $3.3$, $2.7$ and $3.1$ for Washington, New York, California, Florida, Texas and Illinois, respectively.
During the early stage of the outbreak, the $\mathcal{R}_e$ generally has a decreasing trend.
We suspect that  the decline in $\mathcal{R}_e$ may be associated with the implementation of mitigation policies (e.g., statewide stay-at-home orders, shown in Figure \ref{fig:R_eff}) and the increase of public awareness.
Starting from April, the $\mathcal{R}_e$ for these states is maintained around or below 1, indicating (partial) containment of the disease.
However, with the gradual lift of stay-at-home orders and reopening of businesses, we can clearly observe rebounds of $\mathcal{R}_e$ for some states (e.g., Florida) since May.
For all the states, we can observe local fluctuations of  $\mathcal{R}_e$ over time, which may potentially be attributed to some unobserved factors such as social distancing fatigue.
Our analysis is preliminary and does not lead to definitive conclusions about whether a specific intervention is effective in controlling disease spread.
Due to the issue of (potentially unmeasured) confounding, it is very challenging to draw causal inference about the effectiveness of an intervention.
Nevertheless, our analysis can shed light on the transmission dynamics of COVID-19 and may be used as a reference for decision-makers.

\subsection{Test of Fit}
We carry out the Bayesian $\chi^2$ test \citep{johnson2004bayesian} to assess the goodness-of-fit of our model using Illinois data as an example.
First, we choose quantiles $0 \equiv a_0 < a_1 < \cdots < a_{G-1} < a_{G} \equiv 1$, with $p_g = a_g - a_{g-1}$, $g = 1, \ldots, G$.
As suggested by \cite{johnson2004bayesian}, we use $(a_0, \ldots, a_5) = (0, 0.2, 0.4, 0.6, 0.8, 1)$, so $p_g \equiv 0.2$ and $G = 5$.
Next, let $\btheta^{(\ell)}$ be a posterior sample of the model parameters $\btheta$, and let $m_g \left( \btheta^{(\ell)} \right)$ denote the number of observations (i.e., $B_t$'s) such that $\logit \left\{B_t / [ (1 - \alpha^{(\ell)}) I^{U (\ell)}_t] \right\}$ 
falls between the $a_{g-1}$ and $a_{g}$ quantiles of the distribution $\N \left( \by_t^{\top} \bmeta^{(\ell)}, \sigma_{\gamma}^{(\ell)2} \right)$.
Let
\begin{align*}
\omega\left( \btheta^{(\ell)} \right) = \sum_{g = 1}^G \left[ \frac{m_g \left( \btheta^{(\ell)} \right) - (T+1) \cdot p_g }{\sqrt{(T+1) \cdot p_g }} \right]^2.
\end{align*}
Then, under the null hypothesis of a good model fit, the statistic $\omega$ should follow a $\chi^2$-distribution with $G-1 = 4$ degrees of freedom.
A quantile-quantile plot of the posterior samples of $\omega$ against the expected order statistics from a $\chi_4^2$ distribution (Appendix Figure \ref{fig:test_fit}) shows that $\omega$ plausibly comes from a $\chi_4^2$ distribution. 
In addition, we find the proportion of posterior samples of $\omega$ exceeding the 95\% quantile of a $\chi_4^2$ distribution to be 0.053.
There is no evidence of a lack of fit.

\subsection{Forecasts}

\paragraph{Retrospective forecasts} 
As described in Section \ref{sec:predictive}, the proposed method can be used to predict a future observation based on its posterior predictive distribution.
To evaluate the forecasting performance of the proposed model, we conduct within-sample forecasts using Illinois as an example.
Specifically, we split the observations $\bB$ into a training set $\bB^{\tr}$ and a testing set $\bB^{\te}$, where $\bB^{\tr} = (B_0, B_1, \ldots, B_{t^*})$ and $\bB^{\te} = (B_{t^* + 1}, B_{t^* + 2}, \ldots, B_{T})$.
We consider three different scenarios, $t^* \in \{19, 39, 59\}$, so that the training set consists of observations for 20, 40 and 60 days, respectively.
We first sample from the posterior distribution of the parameters evaluated on the training set, $\pi( \btheta \mid \bB^{\tr}, I^D_0)$, and then sample from the posterior predictive distribution of the testing observations, $\pi(\bB^{\te} \mid \bB^{\tr}, I^D_0)$.

\begin{figure}[h!]
\centering
\begin{tabular}{ccc}
\includegraphics[width = 0.31\textwidth]{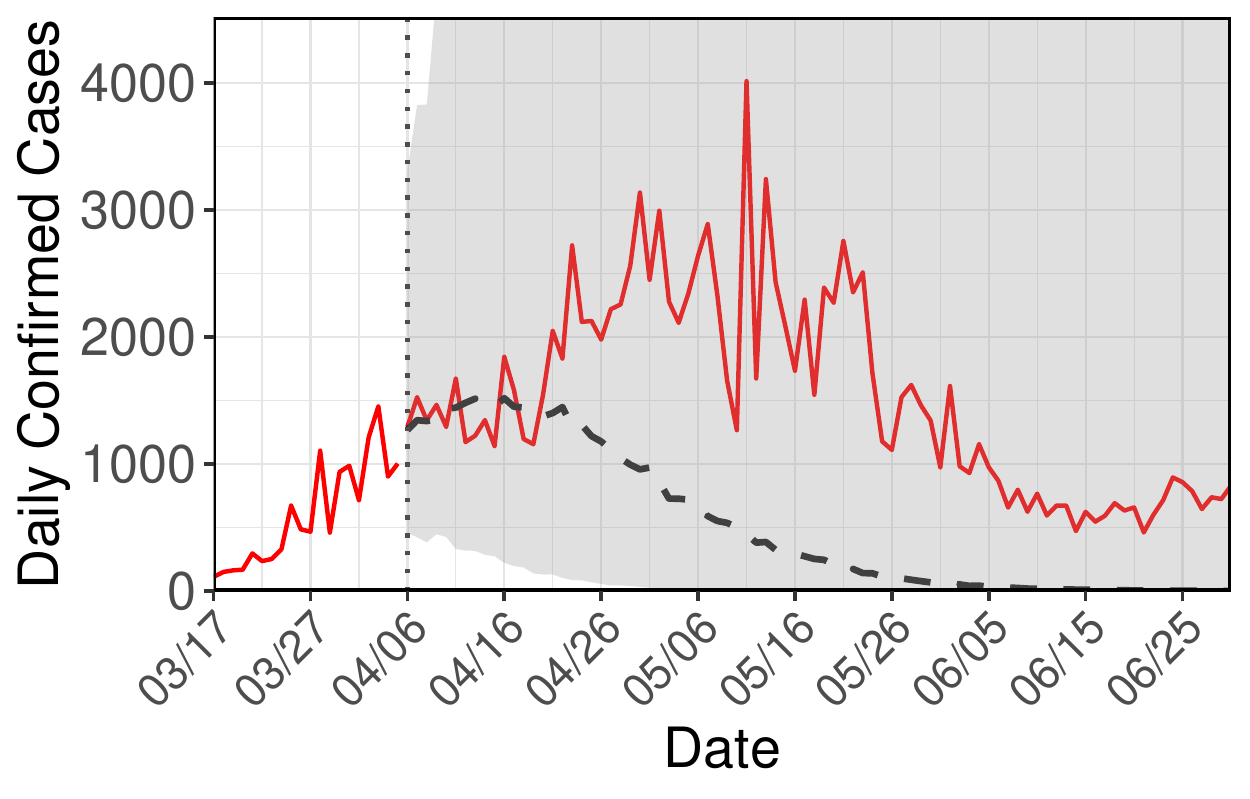} 
&
\includegraphics[width = 0.31\textwidth]{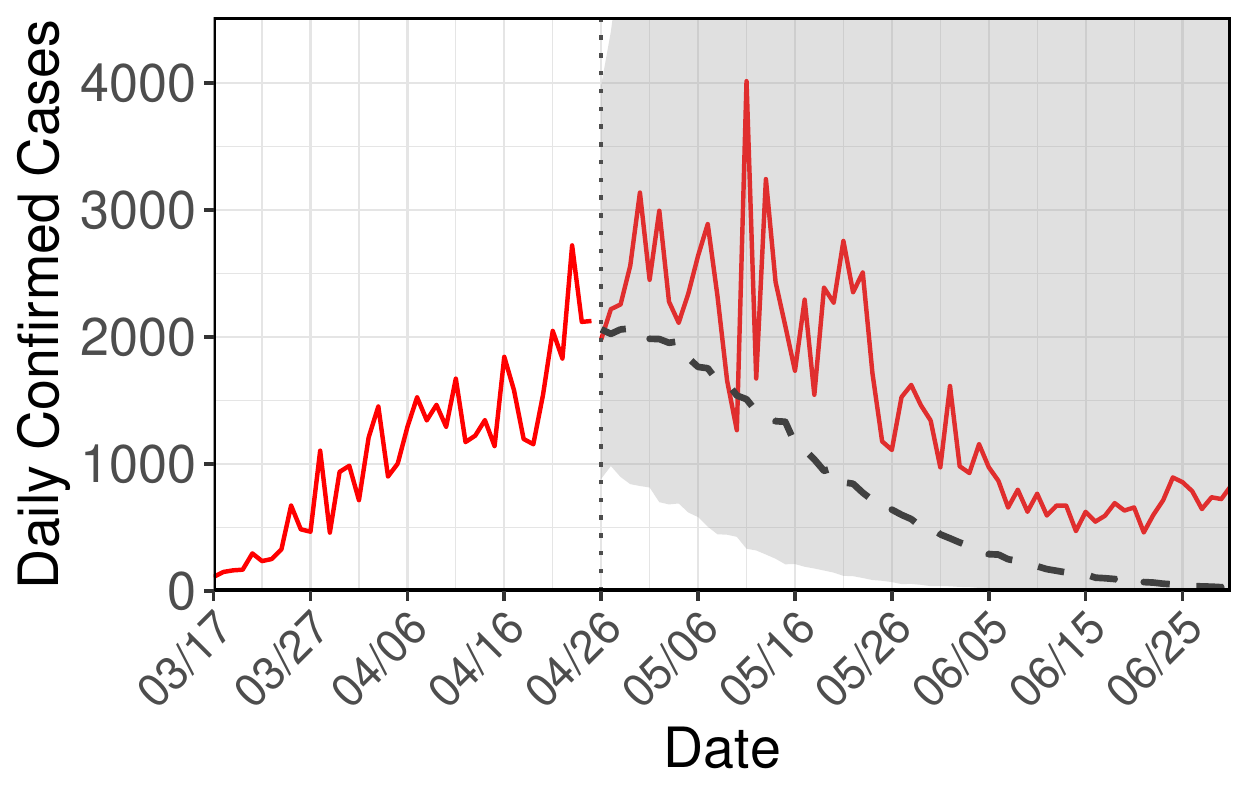} 
&
\includegraphics[width = 0.31\textwidth]{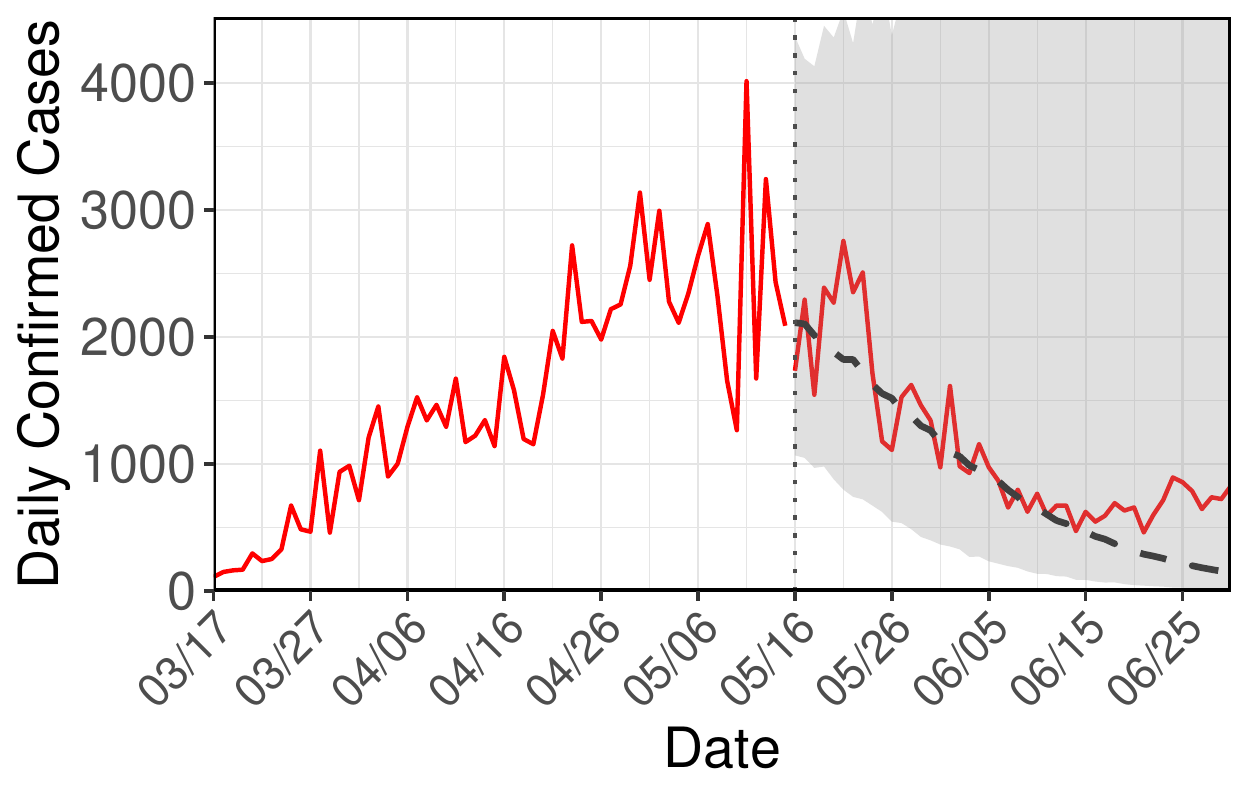}
\\ 
\includegraphics[width = 0.31\textwidth]{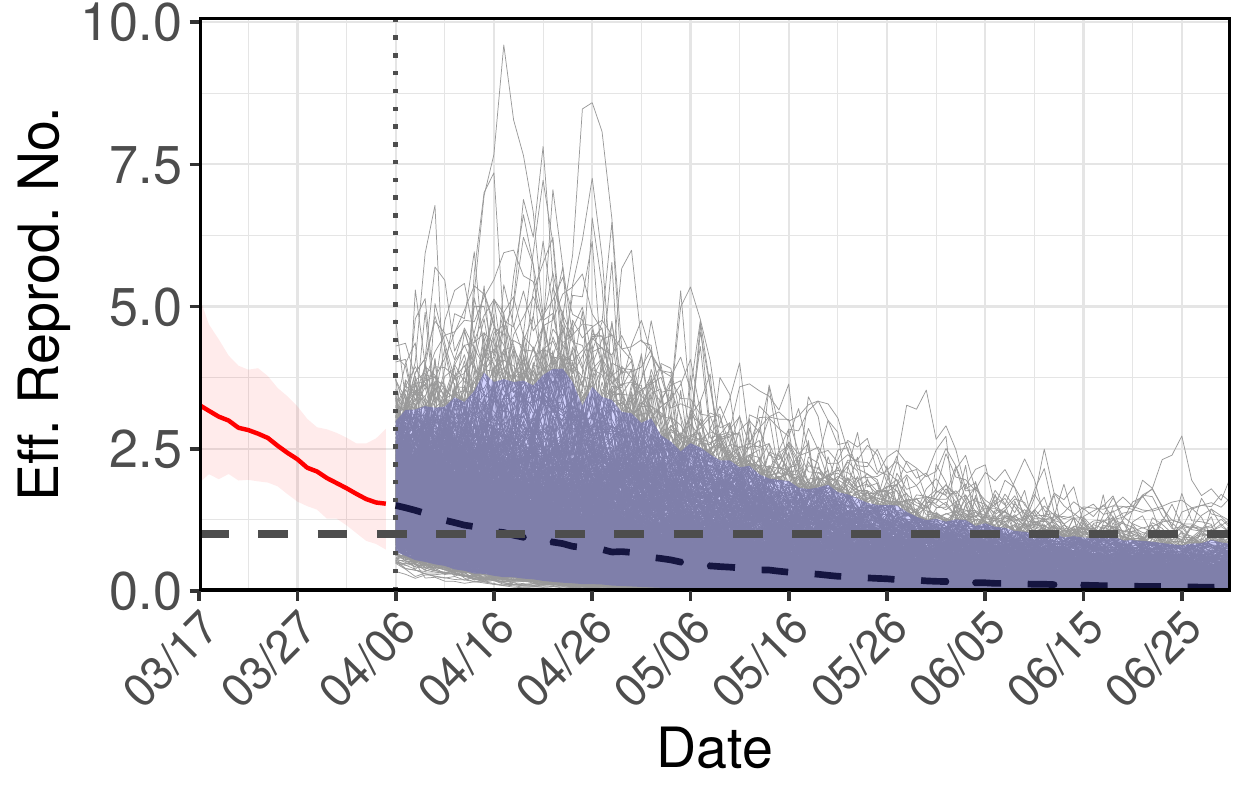} 
&
\includegraphics[width = 0.31\textwidth]{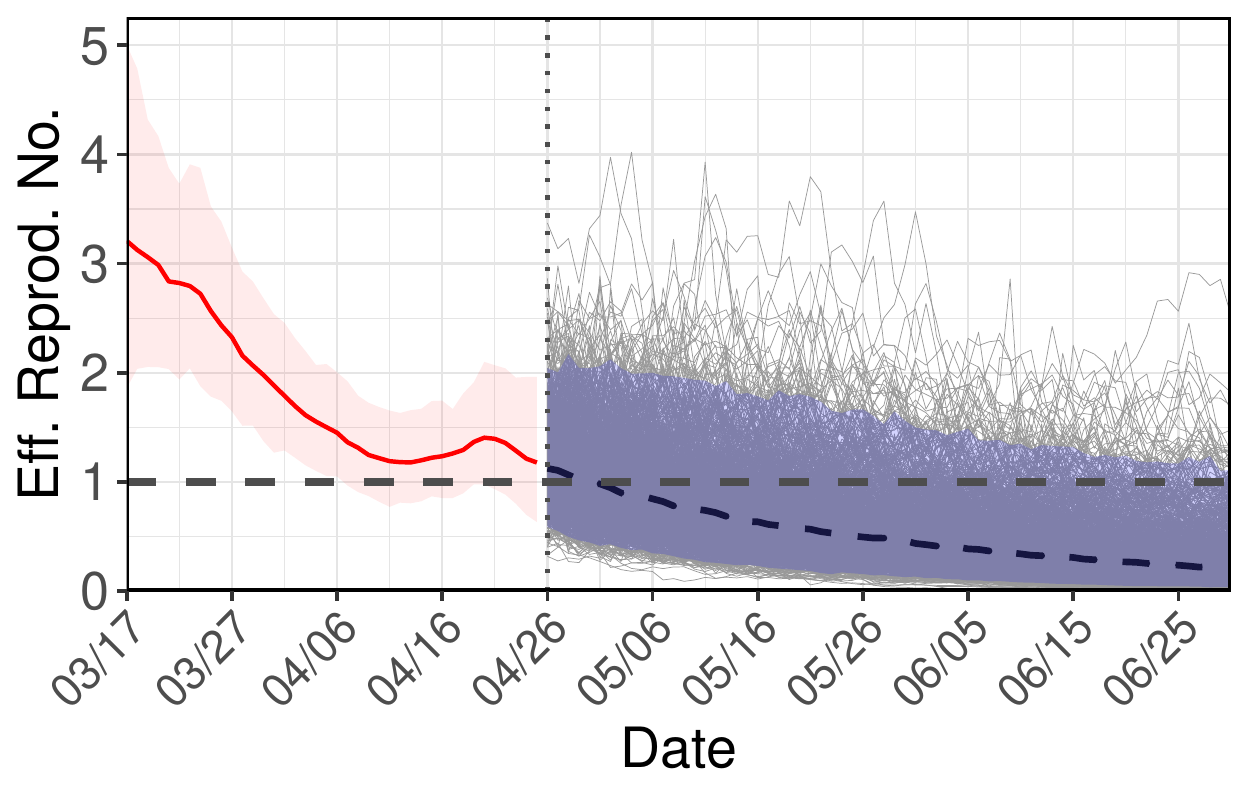} 
&
\includegraphics[width = 0.31\textwidth]{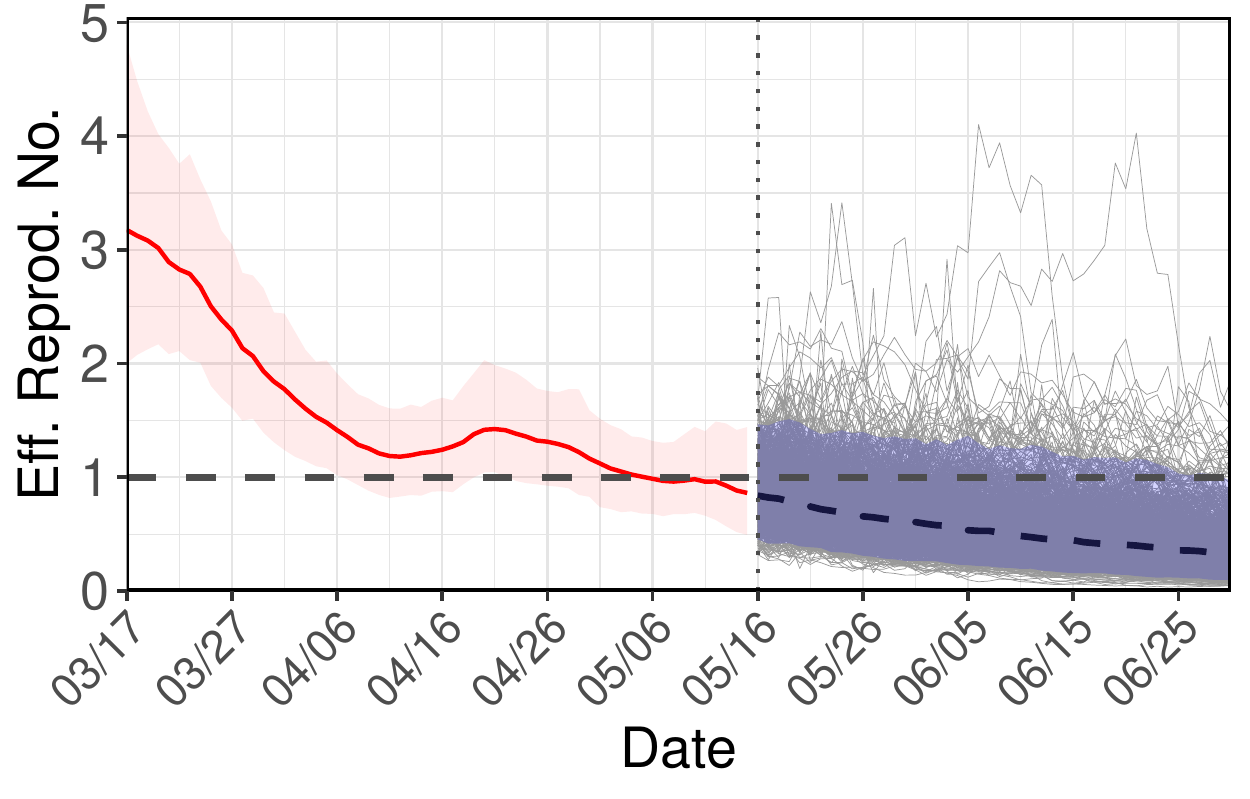} 
\\
(a) 20-day training data &
(b) 40-day training data &
(c) 60-day training data
\end{tabular}
\caption{Within-sample forecasts for Illinois using 20-day, 40-day or 60-day training data. 
The upper panel shows the observed daily confirmed cases (solid red line), and posterior medians (dashed line) and 95\% credible intervals (grey band) for $(\bB^{\te} \mid  \bB^{\tr}, I^D_0)$. The upper bounds of the credible intervals are truncated for better display.
The lower panel shows the posterior medians (solid red line) and 95\% credible intervals (red band) for $[\mathcal{R}_e(0),  \ldots, \mathcal{R}_e(t^*) \mid \bB^{\tr}, I^D_0]$, and posterior medians (dashed line), posterior draws (thin grey lines) and 95\% credible intervals (blue band) for $[\mathcal{R}_e(t^* + 1),  \ldots, \mathcal{R}_e(T) \mid \bB^{\tr}, I^D_0]$.}
\label{fig:within_forecast}
\end{figure}

Figure \ref{fig:within_forecast} shows the forecasting results for the three scenarios.
The 97.5\% percentile of $\pi(\bB^{\te} \mid \bB^{\tr}, I^D_0)$ (i.e., the upper bound of the 95\% credible interval) is truncated in the figure for better display, because it becomes huge with exponential growth.
To better understand the forecasting behavior of the proposed model, the predictions of future $\mathcal{R}_e(t)$'s are also displayed.
Using 20-day training data, the median of $\pi(\bB^{\te} \mid \bB^{\tr}, I^D_0)$ underestimates the actual observations, although the 95\% credible interval covers the observed values.
In general, prediction of an epidemic process is challenging, especially when the epidemiological parameters vary over time.
To see this, notice that there is a rebound of $\mathcal{R}_e(t)$ around April 21, which cannot be captured by the GP prediction rule with 20-day training data.
Since the stay-at-home order is still in effect on April 21, this rebound can neither be captured by policy-related covariates.
To summarize, future predictions are made based on extrapolation of the current trend, and if the trend changes unexpectedly, the predictions will be inaccurate.

With more training data, the prediction accuracy improves, as seen in Figure \ref{fig:within_forecast}(b, c).
Using 60-day training data, the median of $\pi(\bB^{\te} \mid \bB^{\tr}, I^D_0)$ matches well with the actual observations.
Lastly, the short-term predictions (within, say, the next 10 days) are reasonably accurate in all the scenarios.

\begin{figure}[h!]
\centering
\begin{tabular}{cc}
\includegraphics[width = 0.47\textwidth]{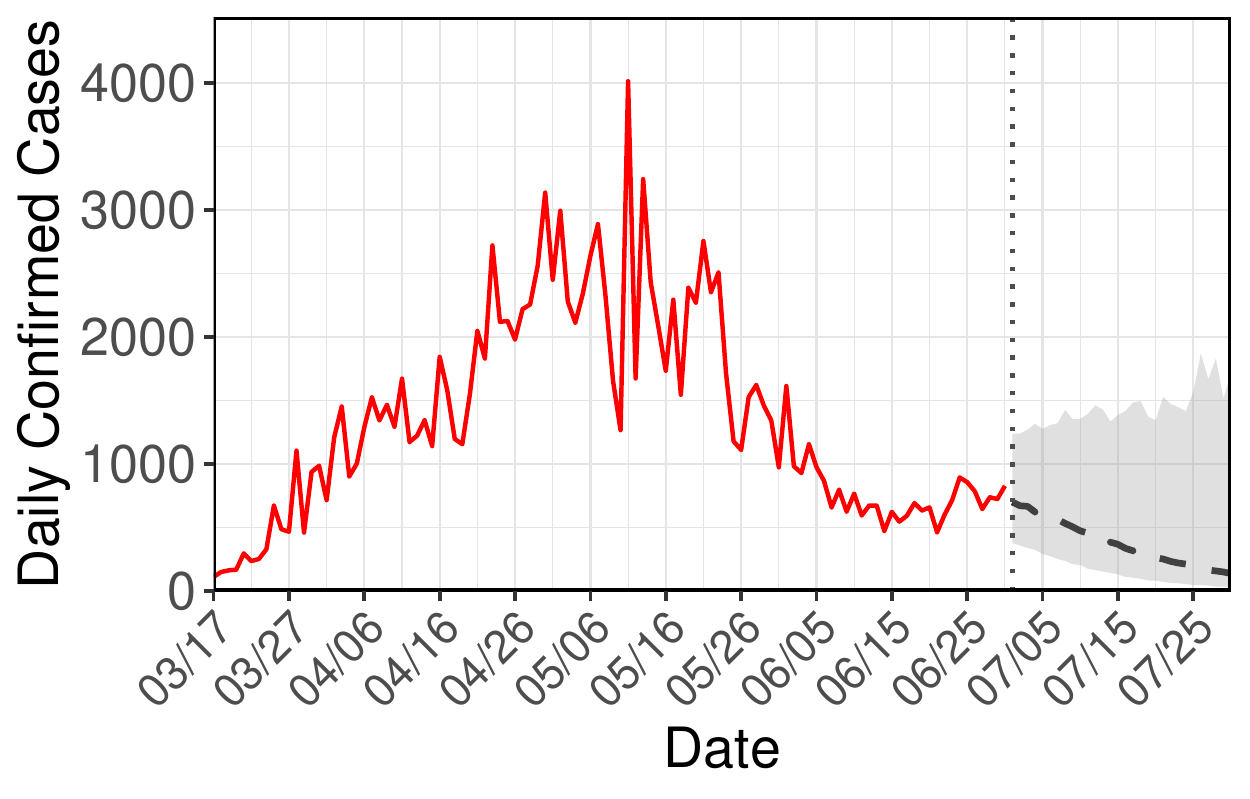}  &
\includegraphics[width = 0.47\textwidth]{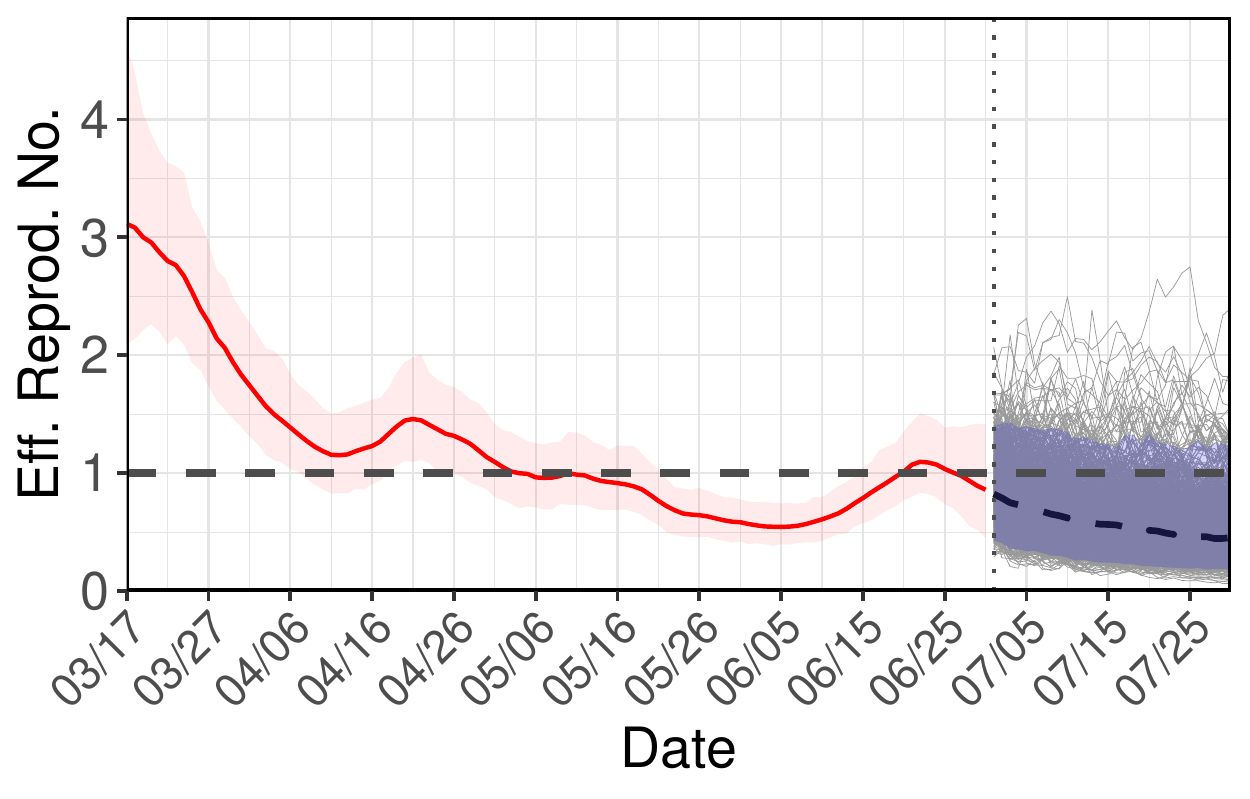} \\
(a) Projected daily confirmed cases & (b) Projected $\mathcal{R}_e(t)$
\end{tabular}
\caption{Out-of-sample forecasts for Illinois in the next 30 days. (a) Observed daily confirmed cases (solid red line), and posterior medians (dashed line) and 95\% credible intervals (grey band) for $(\bB^* \mid \bB, I^D_0)$. (b) Posterior medians (solid red line) and 95\% credible intervals (red band) for $[\mathcal{R}_e(0),  \ldots, \mathcal{R}_e(T) \mid \bB, I^D_0]$, and posterior medians (dashed line), posterior draws (thin grey lines) and 95\% credible intervals (blue band) for $[\mathcal{R}_e(T + 1),  \ldots, \mathcal{R}_e(T + T^*) \mid \bB^{\tr}, I^D_0]$}
\label{fig:out_forecast}
\end{figure}

\paragraph{Prospective forecasts}
To make future predictions, we first sample from $\pi( \btheta \mid \bB, I^D_0)$ and then sample from $\pi(\bB^* \mid \bB, I^D_0)$; recall that $\bB^* = (B_{T+1}, \ldots, B_{T + T^*})$. 
Figure \ref{fig:out_forecast} shows the projected daily confirmed cases and $\mathcal{R}_e(t)$'s for Illinois in the next 30 days (i.e., $T^* = 30$). 
The projections are based on the assumption that the decreasing trend of $\mathcal{R}_e(t)$ continues.
With the lift of the stay-at-home order and the reopening of businesses, it is possible that $\mathcal{R}_e(t)$ will rebound, thus caution is needed in interpreting the forecasting results.

\section{Discussion}
\label{sec:discussion}

We developed a Bayesian approach to statistical inference about the transmission dynamics of COVID-19.
We proposed to estimate the disease transmission rate using GPR, which captures nonlinear and non-monotonic trends without the need of specific parametric assumptions.
A PTMCMC algorithm was used to efficiently sample from the posterior distribution of the epidemiological parameters.
Case studies based on the proposed method revealed the overall decreasing trend of $\mathcal{R}_e$ in six U.S. states (Washington, New York, California, Florida, Texas, and Illinois), which may be associated with the implementation of mitigation policies and the increasing public awareness of the disease.
Projections for future case counts can be made based on extrapolation, although caution is needed in interpreting the forecasting results.

Extensions of the proposed compartmental model can be made in a number of ways. As described in Section \ref{sec:process_model}, it is possible to further split the UI and DI compartments and to incorporate an exposed compartment. We may also split the removed compartment into recovered and deceased compartments.
See, for example, \cite{giordano2020modelling}, \cite{zhang2020prediction} and \cite{aguilar2020investigating}.
Considering more compartments will make the model more realistic.
However, by adding complexity to the current parsimonious model, sampling, estimation and model unidentifiability problems are likely exacerbated \citep{capaldi2012parameter, osthus2017forecasting}.
A possible way out could be to utilize more observable information, such as numbers of recoveries and hospitalizations.
Nevertheless, not every country/region has these data (or accurate measurements of these data) available, and we chose to model only daily confirmed cases to keep our method general enough and applicable to most countries/regions.

The proposed model as in Equation \eqref{eq:covid_model} is a state-space model motivated by a deterministic SIR model. A future direction is to consider a stochastic epidemic model. For example, a model similar to \cite{lekone2006statistical} may be used,
\begin{alignat*}{2}
&S_{t} = S_{t-1} - A_{t-1}, \qquad
&&I^U_{t} = I^U_{t-1}  + A_{t-1} - B_{t-1} - C_{t-1}, \\
&I^D_{t} = I^D_{t-1} + B_{t-1} - D_{t-1}, \qquad
&&R_{t} = R_{t-1} + C_{t-1} + D_{t-1},
\end{alignat*}
where 
\begin{alignat*}{2}
&A_{t} \sim \Bin \left[ S_{t}, 1 - e^{-\beta_{t} ( I^U_{t} + I^D_{t} ) / N } \right], \qquad
&&C_t \sim  \Bin (I^U_{t}, 1 - e^{-\alpha}), \\
&B_{t} \mid C_t \sim \Bin ( I^U_{t} - C_t, 1 -  e^{-\gamma_{t}} ), \qquad
&&D_t \sim \Bin ( I^D_{t}, 1 - e^{-\alpha} ).
\end{alignat*}
Compared to Equation \eqref{eq:covid_model}, this model may better reflect the stochastic nature of the epidemic process. The cost is increased computational complexity.

In our models for the diagnosis rate (Equation \ref{eq:diag_rate}) and transmission rate (Equation \ref{eq:GP_specification}), we allow incorporation of covariates. 
Currently, we only considered an intercept term and a time trend, because empirically we found it hard to identify the effects of other covariates.
Due to Ockham's razor \citep{jefferys1992ockham}, we preferred the simpler model.
More efficient ways to incorporate covariates, potentially based on model selection or variable selection techniques, are worth further investigation.

Our data analysis was carried out separately for each country/region. A nature extension is to model multiple countries/regions jointly using a hierarchical model to achieve borrowing of information, which usually leads to improvements in parameter estimations.
We assumed that the population in each country/region is closed, ignoring immigration and emigration. Arguably, a more realistic model should take into account spatial spread of the disease, as seen in \cite{li2020substantial}.
Again, the main drawbacks to these extensions would be increased computation time.

As discussed in Appendix \ref{supp:sec:identifiability}, the parameters in model \eqref{eq:covid_model} are unidentifiable with only daily confirmed cases ($B_t$) observed, thus parameter estimates are sensitive to prior choices and modeling assumptions.
Many existing models for COVID-19 share the same situation, which could partially explain why different studies may lead to substantially different estimates. For example, some consider the infectious period as the time from infection to recovery or death, which is around 20--30 days \citep{verity2020estimates}. Under this definition, the estimated effective reproduction numbers would be  higher (e.g., \citealp{aguilar2020investigating}).
Therefore, when interpreting the results, it is important to recognize their reliance on underlying assumptions.

Lastly, since the proposed model \eqref{eq:covid_model} is a state-space model, it is of interest to further explore online and sequential algorithms for posterior sampling, such as sequential Monte Carlo \citep{doucet2001introduction, dukic2012tracking}.
In that way, when data at more time points become available, one can update the posterior in an efficient way rather than re-fitting the model to the complete data.

\bibliographystyle{apalike} 
\bibliography{ref_covid19.bib}

\addresseshere

\clearpage

\appendix

\renewcommand\thefigure{\thesection}
\setcounter{figure}{0}  

\section{Parameter Identifiability}
\label{supp:sec:identifiability}

With only daily confirmed cases observed, the parameters in model \eqref{eq:covid_model} are unidentifiable. To see this, consider the following two epidemic processes (indexed by $j = 1$ and $2$),
\begin{align*}
S_{j, t} &= S_{j, t-1} -\beta_{j, t-1} S_{j, t-1} ( I^U_{j, t-1} +  I^D_{j, t-1} ) / N, \\
I^U_{j, t} &= (1 - \alpha_j) I^U_{j, t-1}  + \beta_{j, t-1} S_{j, t-1} ( I^U_{j, t-1} +  I^D_{j, t-1} ) / N - \gamma_{j, t-1} (1 - \alpha_j) I^U_{j, t-1}, \\
I^D_{j, t} &= (1 - \alpha_j) I^D_{j, t-1} + \gamma_{j, t-1} (1 - \alpha_j) I^U_{j, t-1}, \\
R_{j, t} &= R_{j, t-1} + \alpha_j (I^U_{j, t-1} + I^D_{j, t-1}),
\end{align*}
for $t = 1, \ldots, T$.
The observation is the daily increment in confirmed cases, $B_{j, t} = \gamma_{j, t} (1 - \alpha_j) I^U_{j, t}$.
These two processes give rise to identical observations $ B_{1, t}$ and $B_{2, t}$ for all $t$, if
\begin{align}
\gamma_{1, 0} (1 - \alpha_1) I^U_{1, 0} = \gamma_{2, 0} (1 - \alpha_2) I^U_{2, 0}, 
\label{supp:eq:identifiability1}
\end{align}
and 
\begin{multline}
\gamma_{1, t} (1 - \alpha_1) \left[ (1 - \alpha_1) (1 - \gamma_{1, t-1})  I^U_{1, t-1} + 
\beta_{1, t-1} S_{1, t-1} ( I^U_{1, t-1} +  I^D_{1, t-1}) / N \right] = \\
\gamma_{2, t} (1 - \alpha_2) \left[ (1 - \alpha_2) (1 - \gamma_{2, t-1})  I^U_{2, t-1} + 
\beta_{2, t-1} S_{2, t-1} ( I^U_{2, t-1} +  I^D_{2, t-1}) / N \right],
\label{supp:eq:identifiability2}
\end{multline}
for $t = 1, \ldots, T$. In other words, different sets of parameters can lead to exactly the same observed data.
Even if we restrict that  $ (S_{1, 0}, I^U_{1, 0}, I^D_{1, 0} , R_{1, 0}) = (S_{2, 0}, I^U_{2, 0}, I^D_{2, 0} , R_{2, 0})$ (same initial conditions),
$\gamma_{1, t} \equiv \gamma_1$, and $\gamma_{2, t} \equiv \gamma_2$ (constant diagnosis rate), for any $\alpha_1 \neq \alpha_2$ we can still solve Equations \eqref{supp:eq:identifiability1} and \eqref{supp:eq:identifiability2} and get distinct $\{\gamma_1, \beta_{1, t} \}$ and $\{\gamma_2, \beta_{2, t} \}$ that lead to the same observed data.

\begin{figure}[h!]
\begin{center}
\begin{tabular}{ccc}
\includegraphics[page=1, width = 0.31\textwidth]{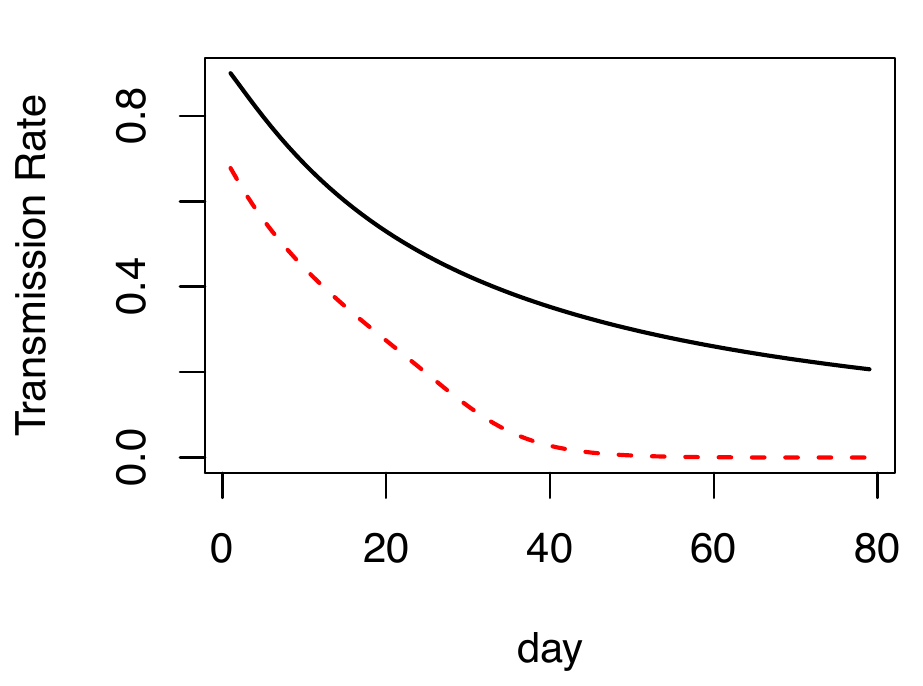} &
\includegraphics[page=2, width = 0.31\textwidth]{figs/identifiability} &
\includegraphics[page=3, width = 0.31\textwidth]{figs/identifiability}   \\
(a) $\beta_{1, t}$ and $\beta_{2, t}$ & (b) $B_{1, t}$ and $B_{2, t}$ (identical) & (c)  $\mathcal{R}_{e}^1 (t)$ and $\mathcal{R}_{e}^2 (t)$
\end{tabular}
\end{center}
\caption{An example of two epidemic processes giving rise to identical observations. Panel (a) shows the distinct transmission rates for the two processes. Panel (b) shows the identical observations given by the two processes. Panel (c) shows the distinct effective reproduction numbers for the two processes.\label{fig:identifiability}}
\end{figure}

A specific example is given below. Consider a population size of $N = 20, 000, 000$.
Suppose there are two epidemic processes with the same initial conditions, $I^U_{1, 0} =  I^U_{2, 0} = 800$, $I^D_{1, 0} =  I^D_{2, 0} = 100$, $R_{1, 0} = R_{2, 0} = 0$, and $S_{1, 0} =  S_{2, 0} = N - 900$.
Suppose further $\alpha_1 = 0.3$, $\alpha_2 = 0.05$,
$\gamma_{1, t} \equiv \gamma_1 = 0.2$, and $\gamma_{2, t} \equiv \gamma_2 = \gamma_1 \cdot (1 - \alpha_1 ) / (1 - \alpha_2) = 0.147$.
Then, the parameters $\beta_{1, t}$ and $\beta_{2, t}$ can be chosen (Figure \ref{fig:identifiability}(a)) such that $\{ B_{1, t} \} $ and $\{ B_{2, t} \}$ are identical (Figure \ref{fig:identifiability}(b)).
The resulting effective reproduction numbers for the two epidemic processes, $\mathcal{R}_{e}^j (t) = (\beta_{j, t} S_{j, t}) / (\alpha_j N)$, are shown in Figure \ref{fig:identifiability}(c) and are quite different.
This example highlights that the parameters in \eqref{eq:covid_model} are unidentifiable in the absence of strong prior information.

\clearpage

\section{Posterior Sampling: Parallel Tempering}
\label{supp:sec:ptmcmc}

\begin{figure}[h!]
\begin{center}
\begin{tabular}{cc}
\includegraphics[page=1, width = 0.45\textwidth]{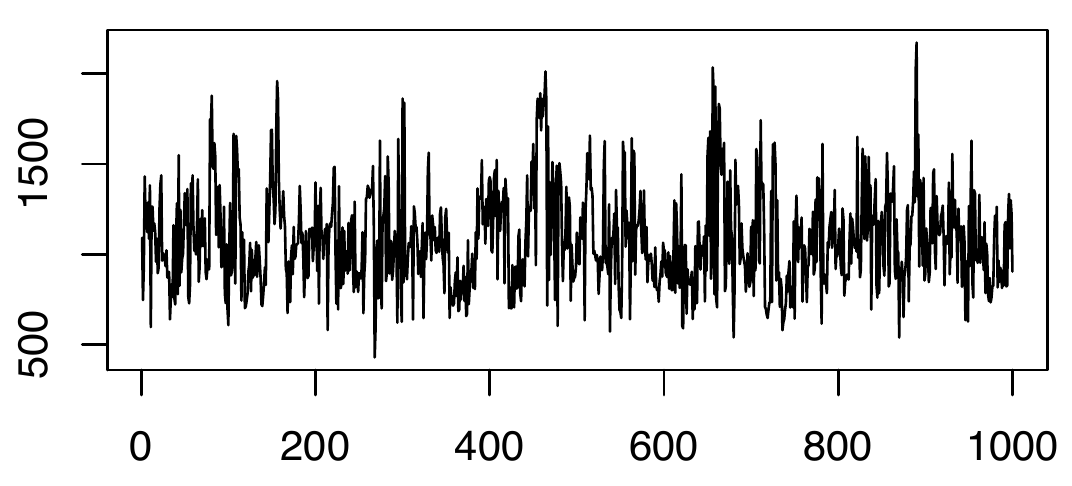} &
\includegraphics[page=2, width = 0.45\textwidth]{figs/PTMCMC} \\
(a) $I^U_0$ using PT, $z_{\text{G}} = 0.38$ &  (b) $I^U_0$ not using PT, $z_{\text{G}} = 4.46$ \\
\includegraphics[page=3, width = 0.45\textwidth]{figs/PTMCMC}  &
\includegraphics[page=4, width = 0.45\textwidth]{figs/PTMCMC} \\
(c) $\eta$ using PT, $z_{\text{G}} = -0.30$ & (d) $\eta$ not using PT, $z_{\text{G}} = -5.90$ \\
\end{tabular}
\end{center}
\caption{Markov chains for $I^U_0$ and $\eta$ using (a, c) or not using (b, d) parallel tempering. 
The posterior correlation of $I^U_0$ and $\eta$ is $-0.82$.
The value $z_{\text{G}}$ refers to Geweke's $z$-score for convergence diagnostic. All chains are based on 50,000 iterations (discarding first 20,000 iterations as burn-in and keeping 1 draw every 30 iterations).\label{fig:parallel_tempering}}
\end{figure}

To demonstrate the advantage of the PT scheme, we show in Figure \ref{fig:parallel_tempering} the Markov chains for $I^U_0$ and $\eta$ generated using or not using PT based on a simulated dataset.
We evaluate the convergence of the chains using Geweke's diagnostic \citep{geweke1991evaluating}.
Under the null hypothesis of chain convergence, Geweke's $z$-score should follow a standard normal distribution.
The $z$-score indicates lack of convergence for the chains generated without PT.

\clearpage

\section{Simulation Studies: Sensitivity Analysis}
\label{supp:sec:sens_analysis}

\begin{figure}[h!]
\centering
\begin{tabular}{cc}
\includegraphics[width = 0.42\textwidth]{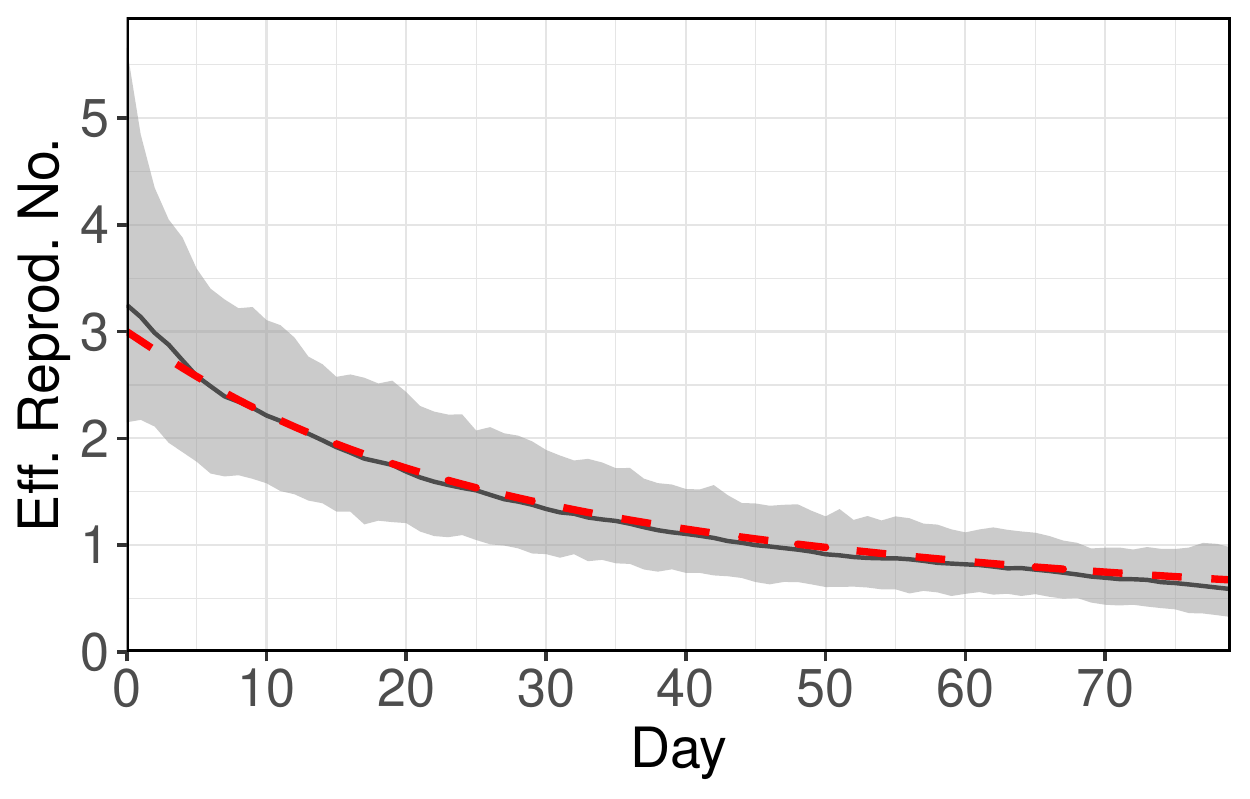}  &
\includegraphics[width = 0.42\textwidth]{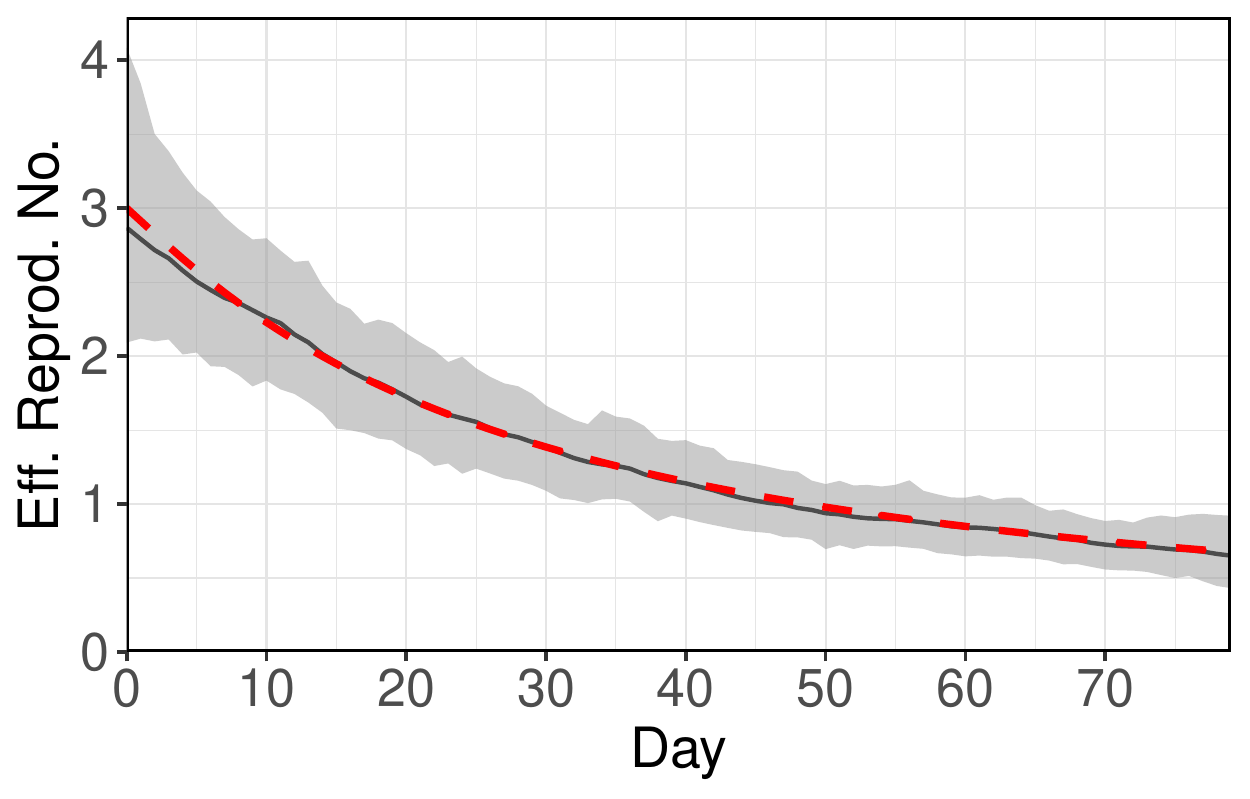} \\
(a) Probit link & (b) Complementary log-log link \\
\includegraphics[width = 0.42\textwidth]{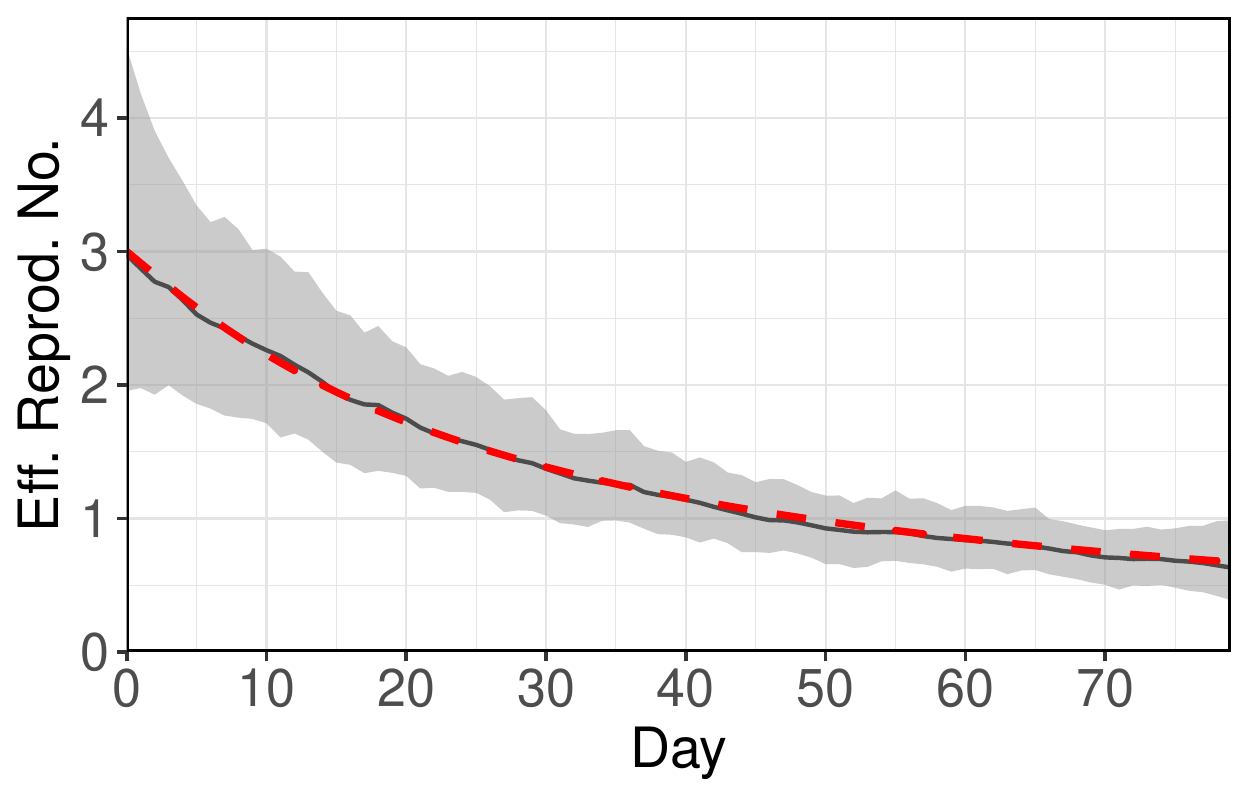} &
\includegraphics[width = 0.42\textwidth]{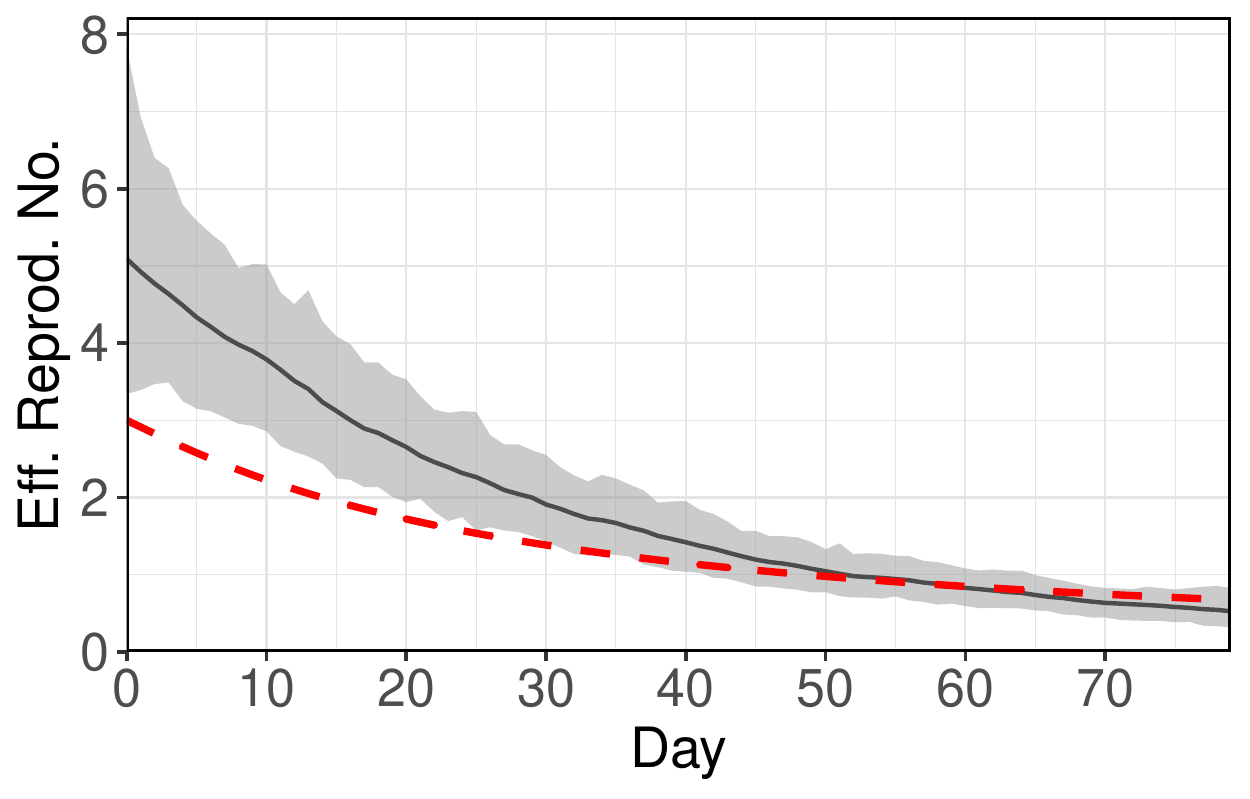}  \\
(c) Larger prior variance for $\alpha^{-1}$ & 
(d) Different prior mean for $\alpha^{-1}$ \\
\end{tabular}
\caption{Simulation Scenario 1. Estimated time-varying effective reproduction numbers (solid black line) with different link functions and priors for $\alpha^{-1}$. The grey band represents the 95\% posterior credible interval, and the dashed red line shows the simulation truth.}
\label{supp:fig:simu_sa}
\end{figure}

We carry out sensitivity analyses to explore how the choice of the link function (Equation \ref{eq:logit_link}) and priors can affect the performance of the proposed method.
We consider the following four settings:
\begin{enumerate}[noitemsep, nolistsep, leftmargin=17mm]
\item[(Set. 1)]  Replacing the default logit link for $\gamma_t$ by the probit link;
\item[(Set. 2)]  Replacing the default logit link for $\gamma_t$ by the complementary log-log link;
\item[(Set. 3)]  Replacing the default prior on $\alpha^{-1}$ by $\alpha^{-1} \sim \Ga(46.5, 5) \mathbbm{1}(\alpha^{-1} \geq 1)$. This leads to a larger prior variance for $\alpha^{-1}$ compared to the default. Recall that the default prior is $\alpha^{-1} \sim \Ga(325.5, 35) \mathbbm{1}(\alpha^{-1} \geq 1)$;
\item[(Set. 4)]  Replacing the default prior on $\alpha^{-1}$ by $\alpha^{-1} \sim \Ga(700, 35) \mathbbm{1}(\alpha^{-1} \geq 1)$. This leads to a different prior mean for $\alpha^{-1}$ compared to the default.
\end{enumerate}
We fit our model to the Simulation Scenario 1 dataset. Figure \ref{supp:fig:simu_sa} shows the estimated time-varying effective reproduction numbers under the four settings.
The estimates are robust to the choice of the link function (Figure \ref{supp:fig:simu_sa}(a, b)).
Also, increasing the prior variance for $\alpha^{-1}$ does not lead to much change in the estimates (Figure \ref{supp:fig:simu_sa}(c)).
Lastly, altering the prior mean for $\alpha^{-1}$ can lead to substantially different estimates (Figure \ref{supp:fig:simu_sa}(d)). This is due to parameter unidentifiability issues (Appendix \ref{supp:sec:identifiability}).
Multiple solutions may explain the observed data equally well, thus the solutions that are more consistent with the prior would be preferred.
Under Setting 4,  the prior for $\alpha^{-1}$ is centered around 20, while the true $\alpha^{-1} = 9.3$.
As a result, the parameter estimates deviate from the simulation truth.

\clearpage

\section{Case Studies: Test of Fit}

\begin{figure}[h!]
\centering
\includegraphics[width = 0.4\textwidth]{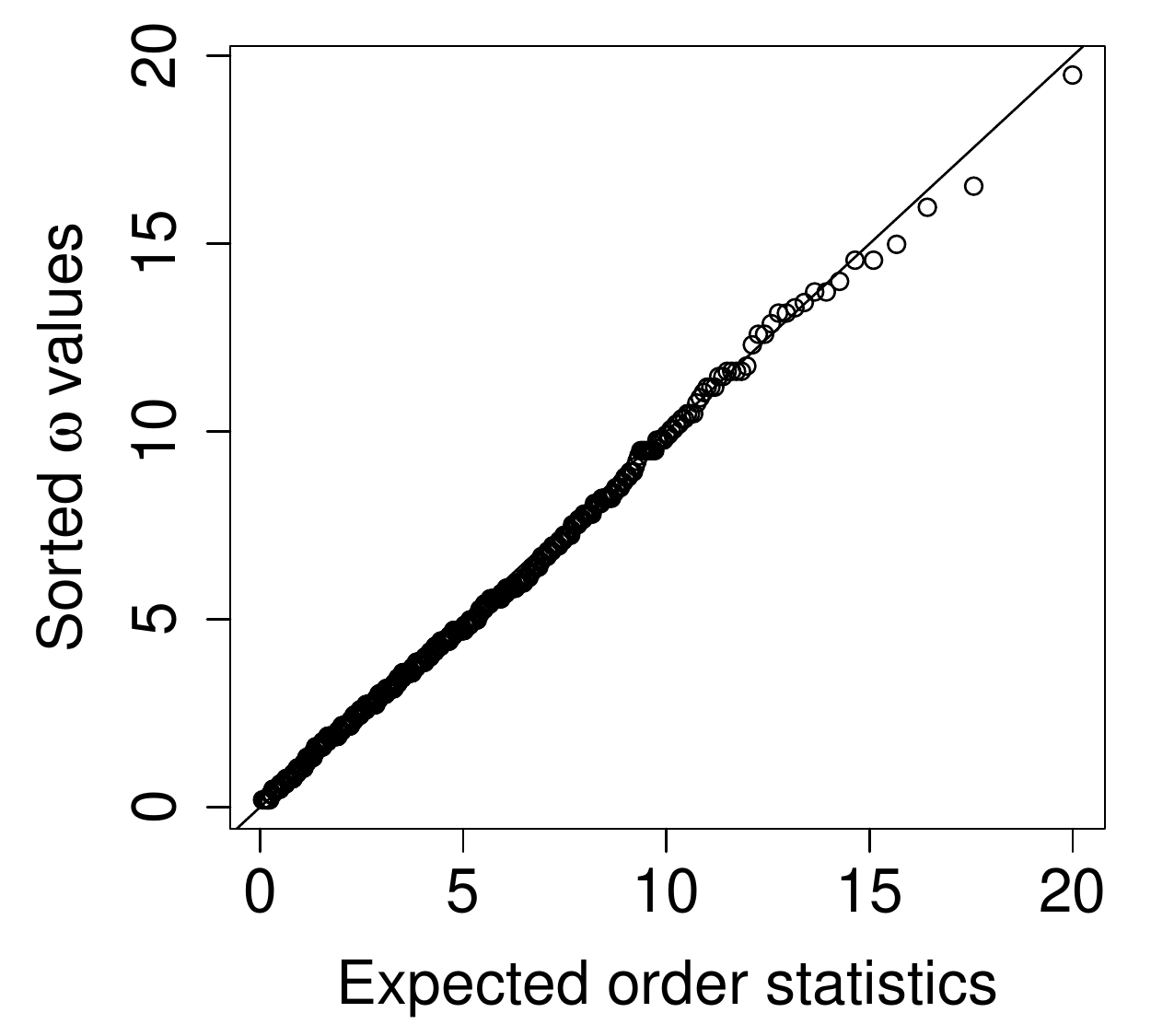} 
\caption{Quantile-quantile plot of posterior samples of the test statistic $\omega$ against expected order statistics from a $\chi_{4}^2$ distribution for the Bayesian $\chi^2$ test}
\label{fig:test_fit}
\end{figure}

We carry out the Bayesian $\chi^2$ test \citep{johnson2004bayesian} to assess the goodness-of-fit of our model using Illinois data as an example.
Under the null hypothesis of a good model fit, the statistic $\omega$ should follow a $\chi_4^2$ distribution.
Figure \ref{fig:test_fit} shows a quantile-quantile plot of posterior samples of $\omega$ against expected order statistics from a $\chi_4^2$ distribution.
There is no evidence that $\omega$ deviates from a $\chi_4^2$ distribution.

\end{document}